\documentclass[a4paper,fleqn,usenatbib]{mnras}
\usepackage{graphicx}
\usepackage[T1]{fontenc}
\usepackage{ae,aecompl}
\usepackage{amssymb,amsmath}
\usepackage{epstopdf}
\usepackage{color}  
\usepackage{float}
\hypersetup{draft} 
\pdfoutput=1

\title[$N$-body simulations and periodic orbits]
{Linking long-term planetary $N$-body simulations with periodic orbits: 
application to white dwarf pollution}
\author[Antoniadou \& Veras]{
Kyriaki I. Antoniadou$^{1}$\thanks{E-mail:kyant@auth.gr},
Dimitri Veras$^{2}$
\\
$^{1}$Department of Physics, Aristotle University of Thessaloniki, 54124 Thessaloniki, Greece
\\
$^{2}$Department of Physics, University of Warwick, Coventry CV4 7AL, UK
}

\date{Accepted XXX. Received YYY; in original form ZZZ}

\pubyear{2016} 

\begin{document}
\label{firstpage}

\pagerange{\pageref{firstpage}--\pageref{lastpage}} 
\maketitle

\begin{abstract}
Mounting discoveries of debris discs orbiting newly-formed stars and white 
dwarfs (WDs) showcase the importance of modeling the long-term evolution of small bodies
in exosystems.  WD debris discs are in particular thought to form from very long-term
(0.1-5.0 Gyr) instability between planets and asteroids. However, the time-consuming nature of 
$N$-body integrators which accurately simulate motion over Gyrs necessitates a judicious 
choice of initial conditions.  The analytical tools known as \textit{periodic orbits} can circumvent 
the guesswork.  Here, we begin a comprehensive analysis directly linking  
periodic orbits with $N$-body integration outcomes with an extensive exploration of the planar
circular restricted three-body problem (CRTBP) with an outer planet and inner asteroid 
near or inside of the $2$:$1$ mean motion resonance.  We run nearly 1000 focused simulations 
for the entire age of the Universe (14 Gyr) 
with initial conditions mapped to the phase space locations surrounding
the unstable and stable periodic orbits
for that commensurability.
In none of our simulations did the planar CRTBP architecture
yield a long-timescale ($\gtrsim 0.25$\% of the age of the Universe) asteroid-star collision.  
The pericentre distance of asteroids which survived beyond this timescale 
($\approx 35$~Myr) varied by at most 
about 60\%. These results help affirm that collisions occur too
quickly to explain WD pollution in the planar CRTBP $2$:$1$ regime, and highlight the
need for further periodic orbit studies with the eccentric and inclined TBP architectures and other
significant orbital period commensurabilities.
\end{abstract}

\begin{keywords}
minor planets, asteroids: general -- Kuiper belt: general -- celestial mechanics 
-- stars: evolution -- stars: white dwarfs -- planets and satellites: dynamical evolution
and stability
\end{keywords}

\section{Introduction}

Unprecedented images of the rings and dust surrounding
HL Tau \citep{almetal2015} provide a glimpse into
the complexity of planet formation.  At the other end
of the life cycle of exosystems, the labile remnant
planetary discs orbiting white dwarfs (WDs) are 
strikingly variable in both brightness and morphology
\citep{wiletal2014,xujur2014,manetal2015,wiletal2015,farihi2016}.
Connecting the future development of systems like HL 
Tau and the past history of WD debris discs
represents an important step towards establishing a
unified evolution theory.

This evolution is not limited to planets.  Both protoplanetary
and WD discs contain dust and potentially asteroids or
other small bodies. In fact, asteroids represent
the favoured progenitors of WD discs for at least three reasons:

\begin{itemize}
  
\item At least one asteroid has now been observed to be disintegrating in real
time around WD 1145+017
\citep{vanetal2015,aloetal2016,ganetal2016,rapetal2016,xuetal2016,zhoetal2016}

\item Planets collide with WDs too infrequently
\citep{veretal2013b,musetal2014,vergae2015,veretal2016} as do exo-Oort cloud comets 
\citep{alcetal1986,veretal2014c,stoetal2015}

\item Measured bulk compositions
in WD atmospheres are incompatible with those from Solar system comets
\citep{zucetal2007,kleetal2010,kleetal2011,gaeetal2012,juretal2012,
xuetal2013,xuetal2014,wiletal2016}

\end{itemize}

Therefore, understanding the interaction between planets and asteroids
is paramount for WD planet studies.  Further, the wide range of
WD cooling ages (the time since the star became a WD) at which
metal pollution is observed (up to 5 Gyr; see \citealt{faretal2011}
and \citealt{koeetal2011}) necessitate understanding the long-term
evolution of planets and asteroids.  For a recent review summarizing
our current knowledge of the long-term behaviour of planetary systems, see \citep{davetal2014},
and for post-main-sequence systems in particular, see \citep{veras2016}.

For the more specific case of an asteroid and planet interacting under the guises of the 
restricted three-body problem (RTBP), a vast body of literature has covered
specific examples and techniques.
For example, \citet{holmur1996} and \citet{murhol1997} analytically and numerically determined 
the long-timescale stability of asteroids in or near mean motion resonance (MMR) under the guise 
of the planar elliptic RTBP, and focused on computing Lyapunov times.  Other measures of chaos, 
such as MEGNO \citep{hinetal2010}, have also been applied to the elliptic RTBP.

$N$-body numerical integrations 
provide a crucial means of determining orbital evolution, but
can be computationally demanding.  Alternatively, predictive analytic 
formulations may yield the desired result more quickly, but their
correctness is subject to validation from $N$-body integrations.
One example is classic Laplace-Lagrange theory 
\citep[see Chapter 7 of][]{murder1999}, which can fail 
to reproduce quantitative behaviour in exosystems when compared to 
$N$-body integrations, even when
the theory is extended to fourth-order in eccentricity \citep{verarm2007}; other
types of extensions yield better results \citep{libsan2013}.
Another example is Lidov-Kozai theory, where differences in its quadrupole-order 
versus octupole-order accuracy are dramatically highlighted through comparison to 
$N$-body integrations \citep[e.g.][]{naoetal2013,naoz2016}. 

Our purpose in this paper is to begin a series of investigations 
comparing long-term $N$-body
integrations to \textit{periodic orbits}, which are defined in Section 2.  
\cite{tsietal2002} also relate numerical integrations to periodic
orbits, but carry out integrations for just 5 Myr and perform a broad sweep
of different period commensurabilities, rather than focusing on one, as
we do here.
Periodic orbits can be treated as analytical tools 
which can gather information regarding a system's dynamics, 
such as in TBPs.
We herein implement the high predictive power of periodic orbits
to explore the 
phase space of a system consisting of a WD, planet and massless
asteroid where the planet and asteroid are evolving in or near
a particular MMR.
Using periodic orbits to remove the guesswork involved in choosing initial 
conditions can reduce the phase space which needs to be explored and increase 
the relevance of the simulation suites that are run.  We provide definitions and
context in Section 2, the results of our numerical simulations in Section 3,
a discussion in Section 4 and our conclusion in Section 5.

\section{Definitions and context}

Now we define a periodic orbit in the planar TBP 
along with other related, relevant terms. 
Throughout this work, we will consider only coplanar systems; lifting off this
restriction would be a catalyst for future studies.  In our TBP, the WD, planet
and asteroid have masses of $m_{\rm WD}$, $m_{\rm P}$ and $m_{\rm A}$, and the
asteroid is initially closer to the WD than the planet.  In Section 2.1, we
define periodic orbits. Then we define the closely-related concept of MMR before linking the two in Section 2.2 and finally focusing on the circular 
RTBP (CRTBP) with a $2$:$1$ internal MMR in Section 2.3.

\subsection{Periodic orbits}\label{po}

Consider a suitable frame of reference that is centred on the centre of mass of the WD
and the planet and rotated, in order for the periodic orbits to be defined and the degrees of freedom to be subsequently reduced through the angular momentum integral \citep[see e.g.][]{hadj12155}.
Degrees of freedom are defined here as
the set $\textbf{Q}(t)$ of positions and velocities of both the planet and asteroid. 
A {\it periodic orbit} is a map $\textbf{Q}(0) = \textbf{Q}(T)$, where the orbit's 
period $T$ satisfies $t = kT$, and $k \ge 1$ is an integer.  Note that $T$ does
not necessarily represent the orbital period of the planet nor the asteroid.

Given the resulting equations of motion, the system can remain invariant 
under certain transformations. As additional restrictions on this system are imposed, 
the number of degrees of freedom decreases. 
For example, the \textit{periodicity conditions} will determine 
if the periodic orbit is classified as \textit{symmetric} or \textit{asymmetric}.
In the planar CRTBP -- where the asteroid is considered massless and the planet
and WD are on circular orbits -- the number 
of degrees of freedom for symmetric and asymmetric periodic orbits is only 2 and
3, respectively \citep[see][]{avk2011}.  

A periodic orbit coincides with a fixed or periodic point on a Poincar\'e surface of section.
This fixed point can be represented by a given set $\textbf{Q}(0)$.  This representation
is the crucial link to initial conditions that supply periodic orbits with predictive power.
Then, through a process known as {\it monoparametric continuation} \citep[see e.g.][]{henon97,hadj12155}, 
these fixed points can be linked together.  The result are smooth curves
known as \textit{characteristic curves}, or \textit{families}, of periodic orbits.
In physical coordinate or orbital element phase space plots, these smooth curves 
act as the visual manifestation of periodic orbits and provide deep insight into the TBP, 
illustrating where a three-body system may be stable, or not, over long timescales.   
In fact, periodic orbits can be classified as ``stable'' 
or ``unstable'' based on a linear stability analysis \citep{marchal90}. An important property of stable symmetric periodic orbits is that the apsidal angle difference\footnote{The apsidal angle difference is the difference
in the arguments of pericentres of the planet and asteroid.} precesses about either $0^{\circ}$ ({\it alignment}) or $180^{\circ}$ ({\it anti-alignment}). It still precesses for asymmetric periodic orbits, but about other angles. The resonant angles librate in a similar manner \cite[for details, see e.g.][]{antoniadou2016}. These angles rotate if the periodic orbits are unstable. The key goal
of this work is to explore how this classification corresponds to the long-term
stability of planet-asteroid systems.

In Hamiltonian systems, stable periodic orbits are surrounded by invariant tori 
in their neighbourhood in phase space, where the motion is regular and quasi-periodic. On the other 
hand, homoclinic webs are formed in the vicinity of the unstable periodic orbits, which trigger 
chaotic motion. In the 
case of weak chaos, the orbits evolve with some irregularity in the 
oscillations of orbital elements and a significant change in the configuration of the system 
is not apparent. If a system is located in strongly chaotic regions, it will eventually 
destabilize, exhibiting collisions or escape. Thus, the long-term stability of a system can be 
guaranteed if it resides in a stability domain in phase space buttressed
by families of stable periodic orbits.

Here, we consider these periodic orbits in the context of long-term $N$-body simulations.

\begin{figure*}
\centerline{
\includegraphics[width=\textwidth]{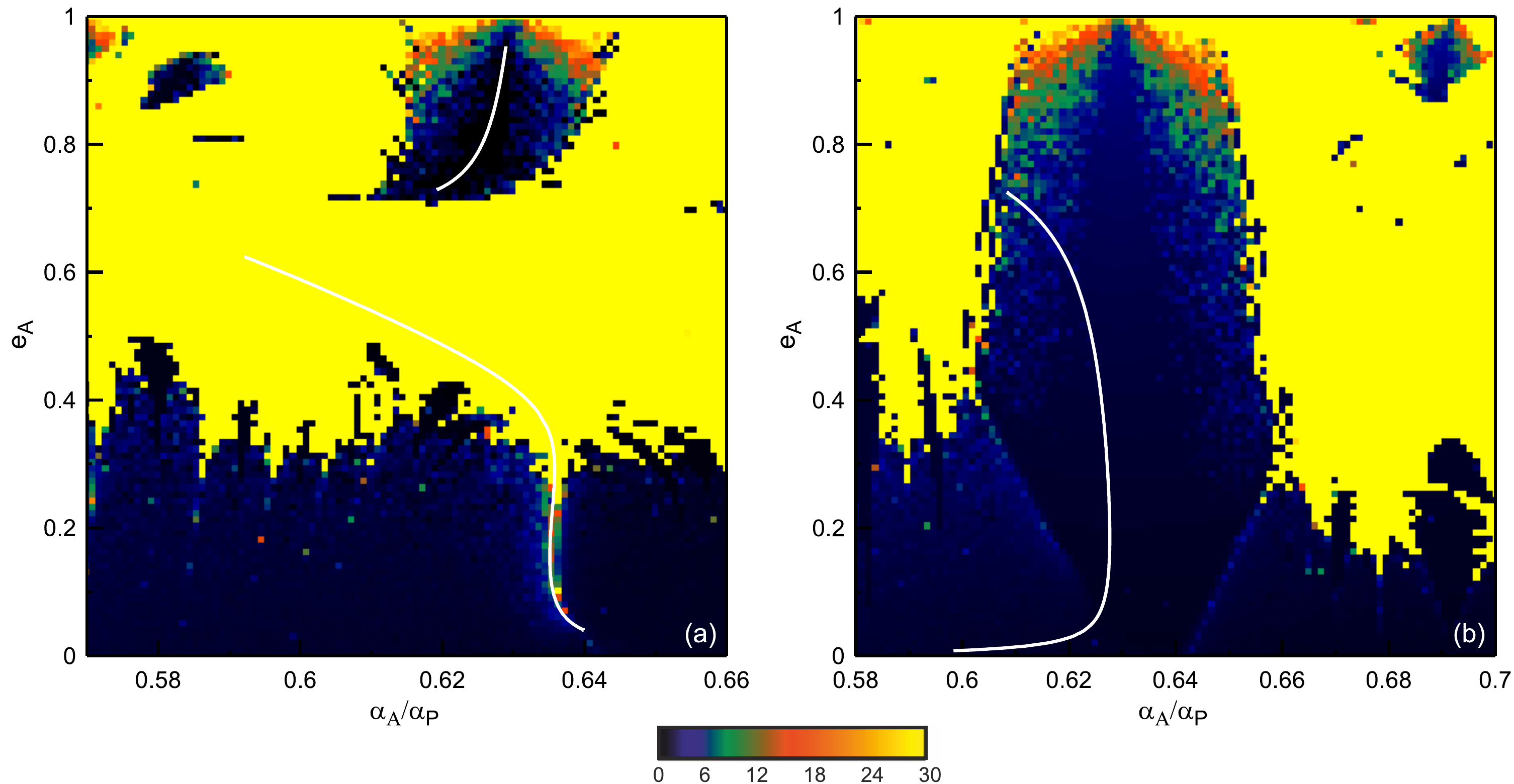}
}
\caption{
Families of periodic orbits (white curves) for the $2$:$1$ 
planar CRTBP where in panel (a) $\omega_{\rm A} = M_{\rm A} = 180^{\circ}$ and in 
panel (b) $\omega_{\rm A} = M_{\rm A} = 0^{\circ}$.  The background colours refer
to the level of order in the phase space of the system by illustrating the logarithmic value of the
Detrended Fast Lyapunov Indicator (DFLI). The families of stable periodic orbits 
in both panels reside in the dark regions, and the family of the unstable ones resides 
in the light region in panel (a).
}
\label{mapsea}
\end{figure*}

\subsection{Link between periodic orbits and resonances (MMRs)}

Because periodic orbits rely on repeating configurations, these orbits are intrinsically
linked to MMRs.  

An MMR is a class of dynamical states.  MMRs have been extensively invoked and studied
\citep[see e.g.][]{var99,lau02,haghi03,voyhadj05,mbf06,avk2011,av2013,av2014,avv2014}.  
The word {\it commensurability} is often
used in a similar context with a similar definition, although the difference between
a simple period ratio and the exact definition of resonance can have significant implications
for the discovery and characterization of exoplanets \citep{agol2005,veretal2011b,verfor2012,nesetal2013,
armetal2015}.

Direct mathematical definitions of MMRs sometimes start
with the time derivative of an angle that appears in a gravitational disturbing function
\citep[e.g. pg. 331 of][]{murder1999} or a Hamiltonian normal form expanded out to a 
finite perturbation order \citep[e.g. pg. 57 of][]{morbidelli2002a}.  
The ability of a \textit{critical} 
(or \textit{resonant}) angle to {\it librate} (or, oscillate) rather than circulate is a hallmark
of many definitions of resonance \citep[see e.g.][]{bfm03,mbf08}. Other definitions incorporate the trajectories in phase
space between separatricies of a given integrable model.  One fundamental
problem with definitions that rely on the expansions (of e.g. potentials in Lagrange's
Planetary Equations) about particular values of orbital elements is that 
the order of the expansion represents a poor metric for accuracy and possibly convergence 
\citep{veras2007}.  One limitation of using librations in definitions of resonance 
is that a libration angle, centre, amplitude and timescale must {\it all} be defined 
\citep{verfor2010}.  Alternatively, a Hamiltonian form defined by just a few
small perturbation terms might necessitate constraining the ranges of orbital elements.

Periodic orbits and MMRs are linked through $T$.  The condition for a periodic
orbit of period T to exist, $\textbf{Q}(0) = \textbf{Q}(T)$, requires that an admissible value
of $T$ is used.  Often, these values are associated to multiples
of the orbital periods of the planet and asteroid such that the ratio of those
multiples corresponds to an orbital period
commensurability $\frac{T_{\rm planet}}{T_{\rm asteroid}}=\frac{p+q}{p}=$ rational, $q$ being the order of the MMR (e.g. $2$:$1$, $3$:$1$).  

Generally families of periodic orbits are generated by \textit{bifurcation points}. However, there exist 
isolated families that do not bifurcate from periodic orbits of other families.
Additionally, some MMRs do not admit particular symmetries 
of periodic orbits. For example, \cite{beau94} and \cite{voy05} indicated that asymmetric 
periodic orbits do not exist in the 
CRTBP in {\it internal resonances} (where the asteroid is the inner body), but rather 
only in {\it external resonances} (where the asteroid is the outer body).  Also, in the
spatial TBP, spatial periodic orbits of a given MMR may correspond to 
{\it inclination-type resonances} \citep{tholis,leeth,vak14,nammor2015} as well as 
the more widely-studied
{\it eccentricity-type resonances}.

\subsection{$2$:$1$ circular restricted three-body problem (CRTBP)}

In this paper, we focus on the families of periodic orbits given
by the internal $2$:$1$ resonance in the planar CRTBP. 
For this commensurability and setup, one can find two
symmetric periodic orbits through which monoparametric continuation will 
yield two different branches of families: one stable branch, and one
unstable branch.  The unstable branch contains two families (see description below), and the stable
branch contains one family, making a total of three families.

The two branches correspond to different orbital architectures, i.e. 
when the asteroid
is at pericentre ($\omega_{\rm A} = 0^{\circ}$) or apocentre ($\omega_{\rm A} = 180^{\circ}$).
In both cases, $M_{\rm A} = \omega_{\rm A}$.  
The variables $M_{\rm A}$ and $\omega_{\rm A}$ are the asteroid's 
initial mean anomaly and argument of pericentre. 
Examples of generation, survival (see Poincar\'e-Birkhoff theorem) 
and continuation of families of periodic orbits from the unperturbed ($m_P=0$) 
to the perturbed ($m_P>0$) restricted case can be found in \citet{boha76},
\citet{h84} and \citet{h93}.

We can visualize these families
with projections on the $(a_{\rm A}/a_{\rm P}, e_{\rm A})$ plane, where $a_{\rm A}$
and $e_{\rm A}$ refer to the initial semimajor axis and eccentricity of the asteroid,
and $a_{\rm P}$ is the initial semimajor axis of the planet.

Figure \ref{mapsea} illustrates these projections when 
$\omega_{\rm A} = M_{\rm A} = 180^{\circ}$
(left panel) and 
$\omega_A = M_{\rm A} = 0^{\circ}$ 
(right panel) 
with white curves.
In Fig. \ref{mapsea}a the branch corresponding to the initial location of the 
asteroid at apocentre is depicted. This branch is divided, via a region of close 
encounters between the planet and the asteroid, into two families 
(separate white curves), which consist 
of unstable symmetric periodic orbits (lower curve) and stable symmetric periodic 
orbits (upper curve). The curve in Fig. \ref{mapsea}b is the branch of family of 
stable symmetric periodic orbits corresponding to the initial location of the 
asteroid at pericentre.

None of these curves can be fit with simple empirical formulae, and their seemingly-arbitrary 
nature hint at the complexity of the planar $2$:$1$ CRTBP.  For example, the 
curves never reach $e_{\rm A} = 0$; a gap is formed during continuation from the unperturbed to perturbed case, because of the order of the resonance.  For additional details, please see e.g., 
\cite{voyhadj05}, \cite{hadj06} and \cite{vkh09}.

\subsubsection{Detrended Fast Lyapunov Indicator (DFLI)}

\begin{figure*}
\centerline{
\includegraphics[height=7.8cm]{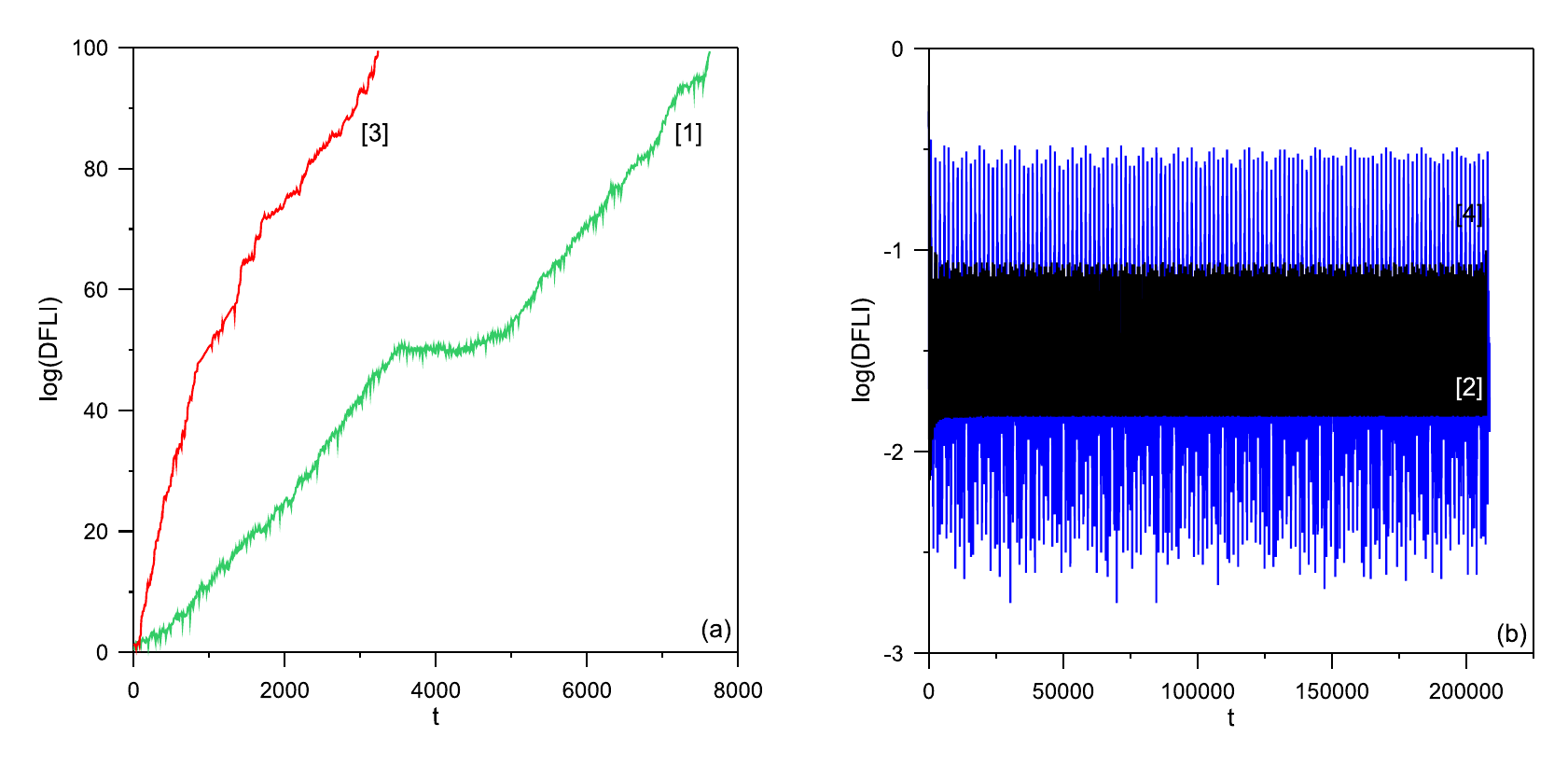}
}
\caption{
Evolution of the Detrended Fast Lyapunov Indicator (DFLI) for four particular systems: 
[1] {\it green line, panel a}: centred on an unstable periodic orbit from Fig. \ref{mapsea}a 
    corresponding to ($a_{\rm A}/a_{\rm P}=0.62498102$, $e_A=0.4563692$, and $\omega_{\rm A} = M_{\rm A} = 180^{\circ}$),
[2] {\it black line, panel b}: centred on a stable periodic orbit from Fig. \ref{mapsea}a 
    corresponding to ($a_{\rm A}/a_{\rm P}=0.62499828$, $e_A=0.7874083$, and $\omega_{\rm A} = M_{\rm A} = 180^{\circ}$),
[3] {\it red line, panel a}: in the vicinity of an unstable periodic orbit, 
    corresponding to ($a_{\rm A}/a_{\rm P}=0.625200$, $e_A=0.629990$, and $\omega_{\rm A} = M_{\rm A} = 180^{\circ}$),
[4] {\it blue line, panel b}: in the vicinity of a stable periodic orbit, 
    corresponding to ($a_{\rm A}/a_{\rm P}=0.625200$, $e_A=0.873470$, and $\omega_{\rm A} = M_{\rm A} = 0^{\circ}$).
The DFLI suggests that the first and third systems are irregular or chaotic, and that the second and fourth
are regular, or stable.
}
\label{dflievol}
\end{figure*}

Plotted in the colourful background behind the lines are contours which indicate the Lyapunov
time, a measure of chaos in the system.  Specifically, the colour of each point of the plane indicates 
the logarithmic values of the Detrended Fast Lyapunov Indicator (DFLI) (FLI divided by t) \citep{froe97,voyatzis08} defined as 
$$DFLI(t)=\frac{1}{t}max\left\{\left|\xi_1 (t)\right|,\left|\xi_2 (t)\right|\right\},$$
where $\xi_i$ are the deviation vectors of the orbit (initially orthogonal) computed after numerical integration of the variational equations of the system for $t_{\rm max} = 2.5 \times 10^5$ years for each initial condition. 
These contours are not the result of full $N$-body integrations, but rather provide a quick
and efficient representation of the phase space. For regular orbits this 
indicator tends to a constant value, whereas for irregular orbits the indicator increases exponentially 
over time taking very large values (see Fig. \ref{dflievol} for a demonstration). On the DFLI maps in Fig. \ref{mapsea} and thereafter, 
dark-coloured regions indicate regular orbits, whereas regions of pale colour represent the the ones where chaotic nature is detected. The integration of the systems was terminated when the index exceeded the value $10^{30}$ and the orbit was accordingly classified (see also the coloured bar in Fig. \ref{mapsea}).  

DFLI maps are not direct substitutes for $N$-body numerical integrations, and in particular
long-term $N$-body numerical integrations.  Like periodic orbits, DFLI maps represent tools
that can help motivate the initial conditions for $N$-body simulations.

The correspondence between the DFLI maps and periodic orbits themselves in both 
panels of Fig. \ref{mapsea} is robust. Phase space is \textit{built} around the periodic orbits given their linear stability. The families of stable periodic orbits are 
located entirely within dark-coloured regions and constitute the backbone of stability domains, and the families of unstable periodic orbits 
are located entirely within pale-coloured regions \citep[see also][]{av16}.

\begin{figure}
\centerline{
\includegraphics[height=6.5cm]{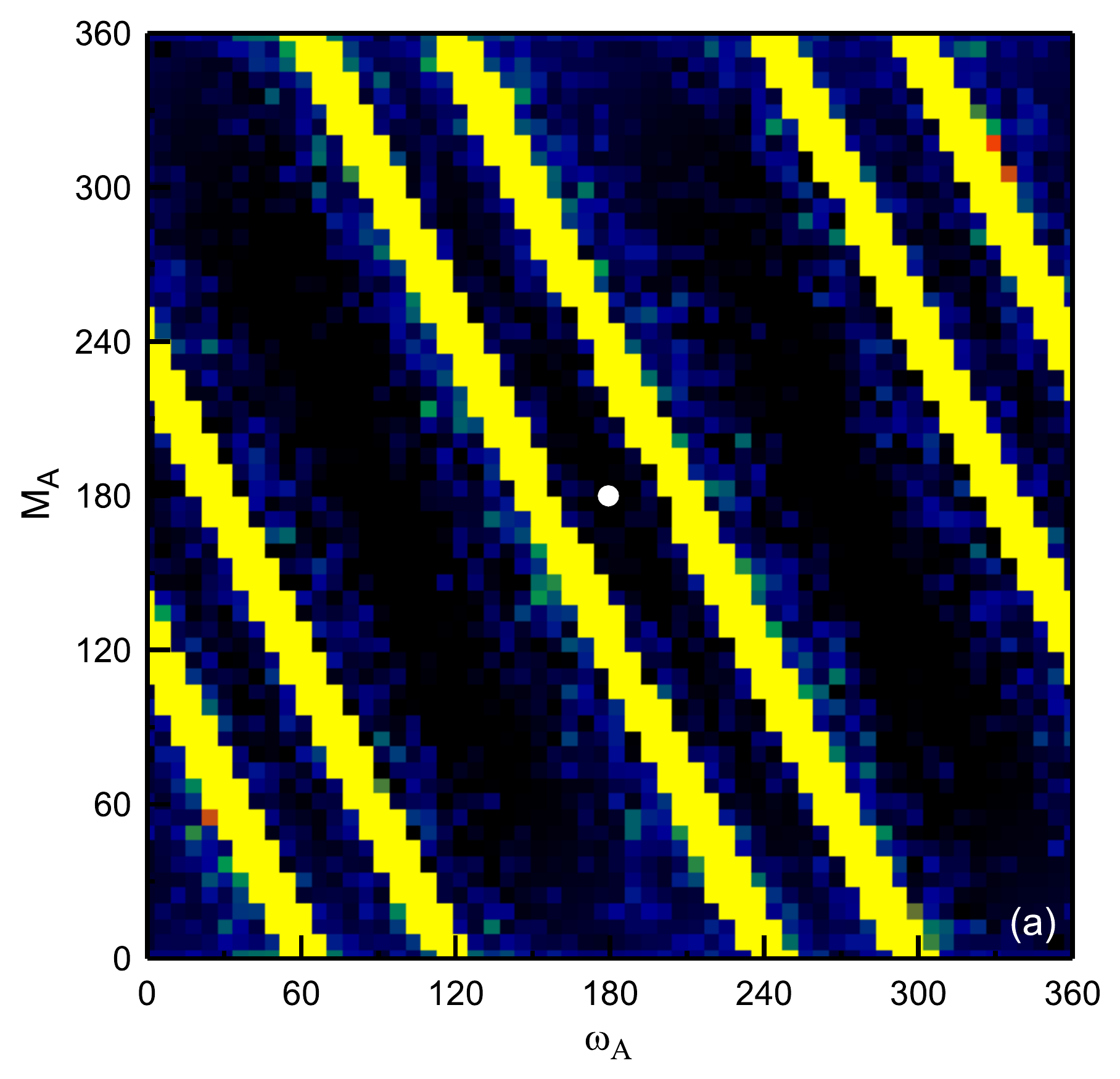}
}
\centerline{
\includegraphics[height=6.5cm]{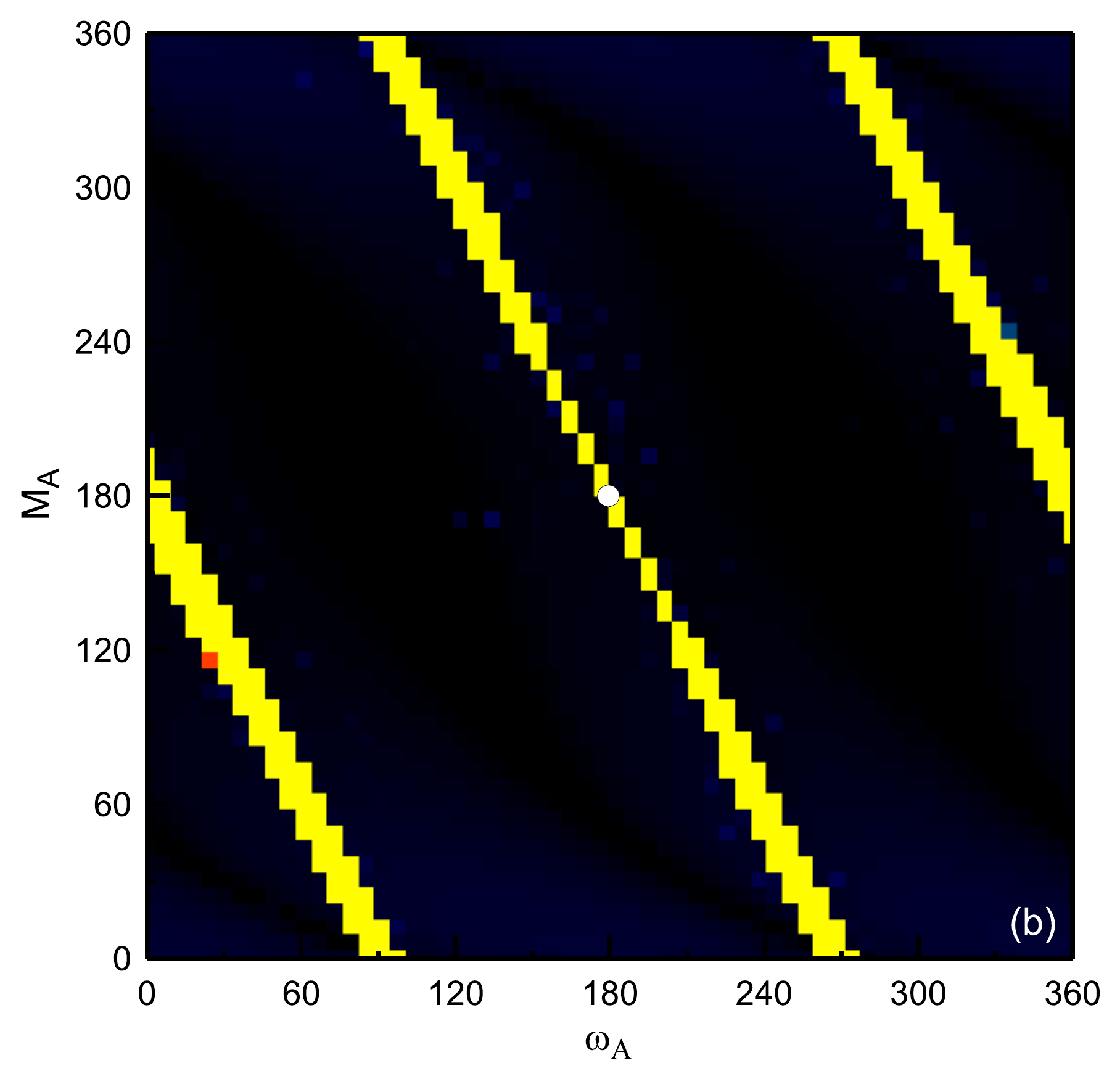}
}
\centerline{
\includegraphics[height=6.5cm]{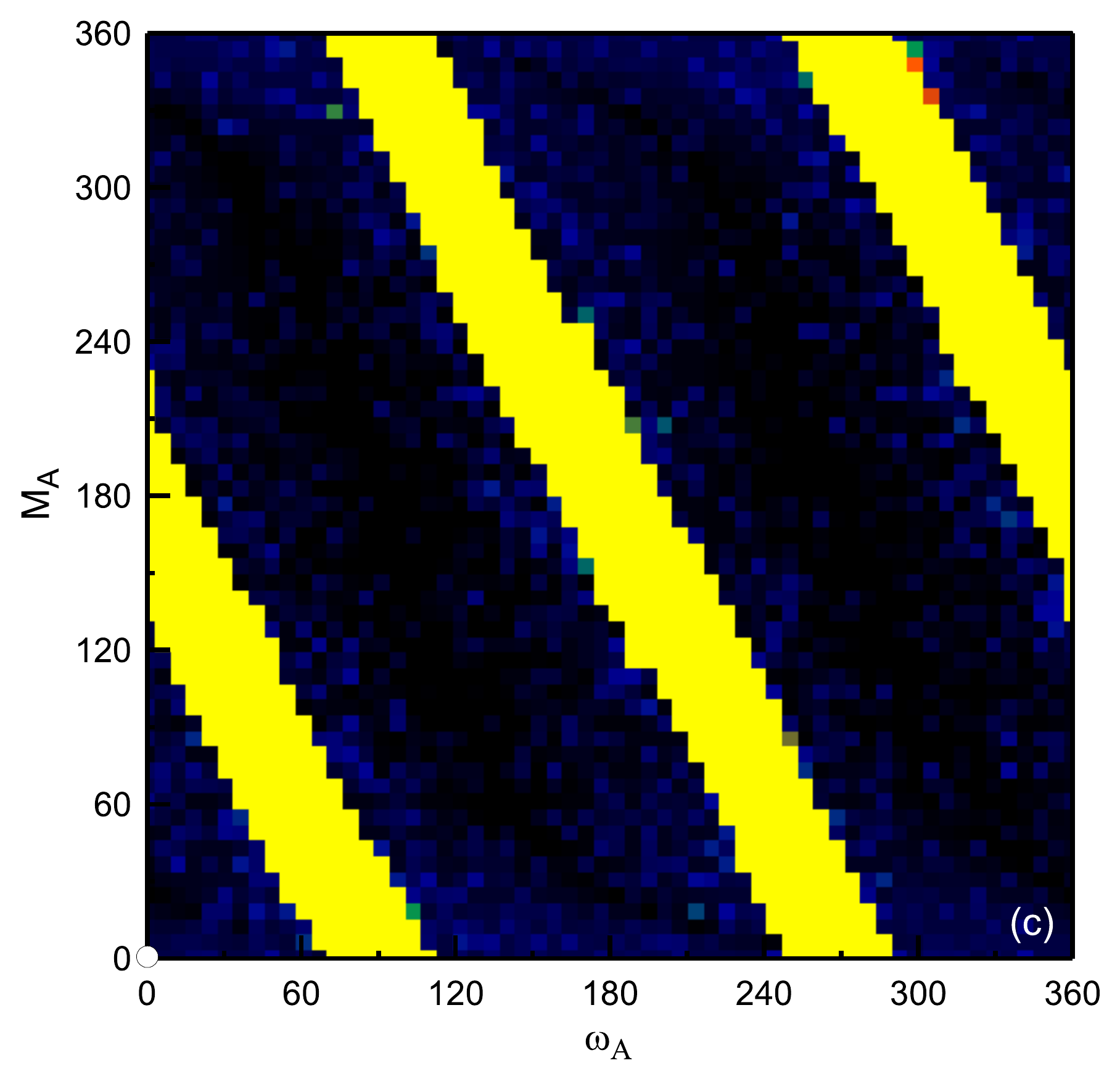}
}
\caption{
DFLI maps on plane $(\omega_{\rm A}, M_{\rm A})$ for 
({\it panel a, with $a_{\rm A}/a_{\rm P}=0.627517$ and $e_{\rm A}=0.864125$}):
a system which intersects with a stable periodic orbit
at $\omega_{\rm A}=M_{\rm A} = 180^{\circ}$,
({\it panel b, with $a_{\rm A}/a_{\rm P}=0.634690$ and $e_{\rm A}=0.358225$}):
a system which intersects with an unstable periodic orbit at 
$\omega_{\rm A}=M_{\rm A} = 180^{\circ}$, and
({\it panel c, with $a_{\rm A}/a_{\rm P}=0.620572$ and $e_{\rm A}=0.600409$}):
a system which intersects with a stable periodic orbit at 
$\omega_{\rm A}=M_{\rm A} = 0^{\circ}$. The periodic orbits are depicted by white circles. These plots reveal how regions of chaotic evolution can be tightly constrained 
in narrow strips in phase space.
}
\label{mapsapowMMOD}
\end{figure}

\begin{figure}
\centerline{
\includegraphics[height=6.5cm]{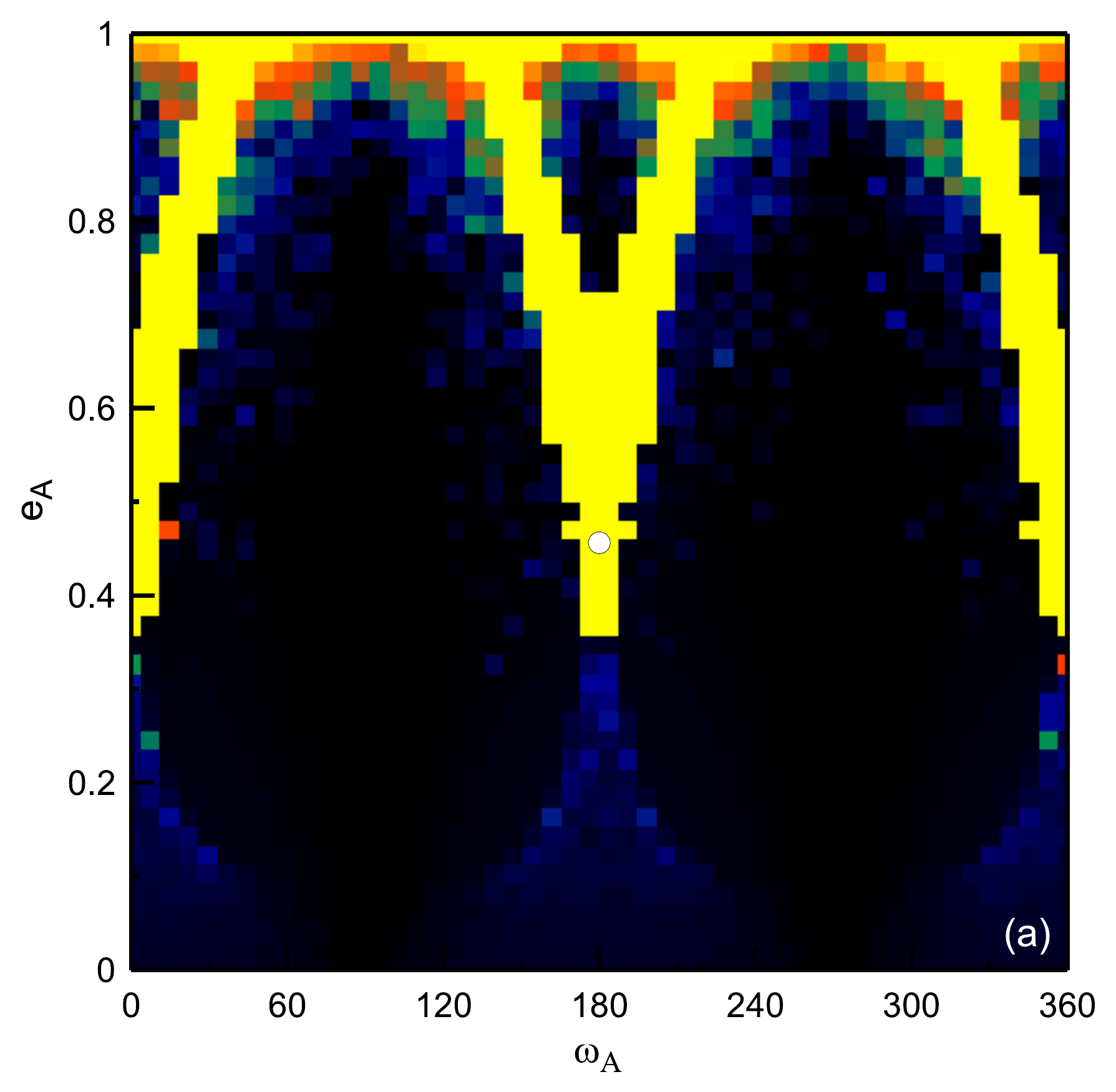}
}
\centerline{}
\centerline{
\includegraphics[height=6.5cm]{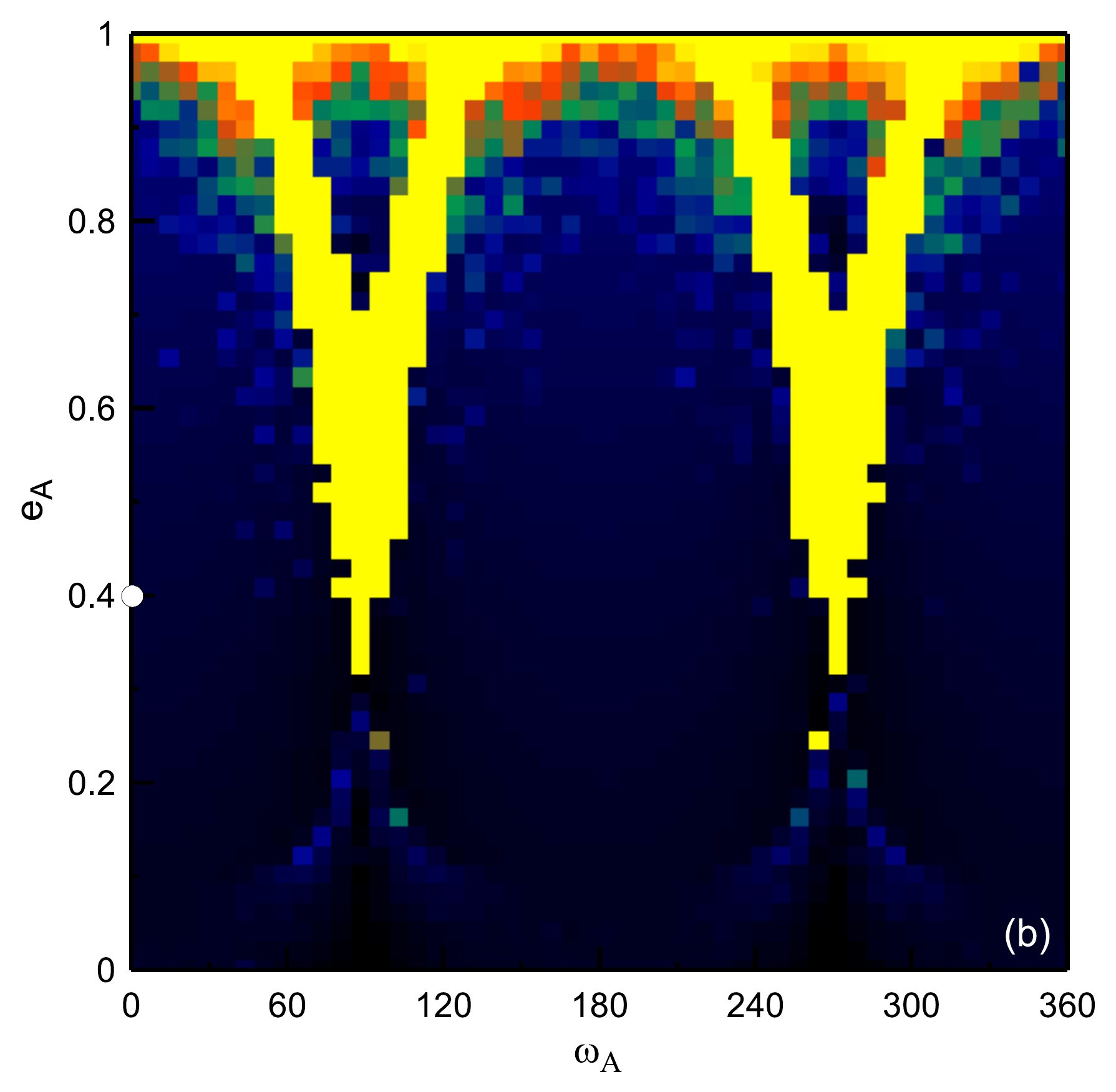}
}
\caption{
DFLI maps on plane $(\omega_{\rm A}, e_{\rm A})$ for 
({\it panel a, with $a_{\rm A}/a_{\rm P}= 0.625603$ and $M_{\rm A}=180^{\circ}$}):
a system which intersects with an unstable periodic orbit at
$e_{\rm A} = 0.456369$, $\omega_{\rm A} = 180^{\circ}$, and
({\it panel b, with $a_{\rm A}/a_{\rm P}= 0.626144$ and $M_{\rm A}=0^{\circ}$}):
a system which intersects with a stable periodic orbit at
$e_{\rm A} = 0.399388$, $\omega_{\rm A} = 0^{\circ}$. The periodic orbits are depicted by white circles. These plots reveal the oscillatory shapes of the strips in this
phase space in which chaotic motion was detected.
}
\label{mapsapoperweMOD}
\end{figure}

\subsubsection{Unraveling phase space}

In order to decipher phase space we additionally construct maps on different planes. 

In Fig. \ref{mapsapowMMOD}, we keep fixed ($a_{\rm A}/a_{\rm P}, e_{\rm A}$), which are provided in the caption, and vary $M_{\rm A}$ and $\omega_{\rm A}$. In panel a, we keep fixed the values of a stable periodic orbit of the stable family of Fig. \ref{mapsea}a and as a result, the regular regions are built around the  $(\omega_{\rm A},M_{\rm A})=(180^\circ,180^\circ)$ and its equivalent\footnote{Due to the symmetry and the MMR, different initial locations ($M_{\rm A}$ or $\omega_{\rm A}$) of the asteroid (the initial location of the planet on the circular orbit is considered the same) on the elliptic orbit at $t=0$ and $t=T/2$ are equivalent in pairs.} configurations. But, one can see in Fig. \ref{mapsea}b that this point falls also within the regular region of the stable family of that configuration, i.e. when $(\omega_{\rm A},M_{\rm A})=(0^\circ,0^\circ)$. Hence, in Fig. \ref{mapsapowMMOD}a we can also observe the regular regions shaped around the latter configuration and its equivalent ones.   In Fig. \ref{mapsapowMMOD}b, we have chosen an unstable periodic orbit of the unstable family of Fig. \ref{mapsea}a which as above falls also within the region of stability in Fig. \ref{mapsea}b. Hence, we can observe both the very thin (keep in mind that the eccentricity value is low) region of irregular orbits existing in the configuration $(\omega_{\rm A},M_{\rm A})=(180^\circ,180^\circ)$ and the regions of regular motion for the configuration $(\omega_{\rm A},M_{\rm A})=(0^\circ,0^\circ)$. Again, the repetitive strips of each behaviour are due to the MMR. In Fig. \ref{mapsapowMMOD}c, we have chosen an unstable periodic orbit of the family in Fig. \ref{mapsea}a for a greater value of the eccentricity. Thus, in comparison with Fig. \ref{mapsapowMMOD}b the regions where irregular motion was detected are larger.

In Fig. \ref{mapsapoperweMOD} we keep fixed ($a_{\rm A}/a_{\rm P}, M_{\rm A}$) and vary $e_{\rm A}$ and $\omega_{\rm A}$ for an unstable (panel a) and stable (panel b) periodic orbit, whose orbital elements are provided in the caption. It is a straightforward observation that both of the structures of regular and irregular orbits seen in Fig. \ref{mapsea} for the two different configurations, i.e. when $(\omega_{\rm A},M_{\rm A})=(180^\circ,180^\circ)$ and $(\omega_{\rm A},M_{\rm A})=(0^\circ,0^\circ)$ are apparent and being repeated for their equivalent configurations given the $M_{\rm A}$ here, as well. The fixed semimajor axes are almost the same for both panels, whereas the eccentricity in panel a is higher. Thus, we can observe how the regions where irregular evolution was detected via the DFLI are altered.

\section{N-body integration setup}

Part of the challenge of relating $N$-body simulations to periodic orbits
is that the former does not map bijectively to the latter.  $N$-body simulations
require more input parameters than degrees of freedom in periodic orbits.  
Also, the development of periodic orbits
has historically relied on normalized coordinates, such that, 
for example, the Gravitational constant and total system mass are both set to unity.

\subsection{Scaling factors}

The families of periodic orbits have been computed under the
assumption that the total mass of the system equals unity, $m_{\rm Star}+m_{\rm P}=1$, where $m_{\rm P} = 10^{-3}$.  We
have to transform this scaling into real units.
The invariance of the equations of motion for periodic orbits for different
scalings is described in Section 3.2.1.1 of \cite{antoniadou2014}.  This analysis
reveals that the appropriate scaling factor $\zeta$ is
\begin{eqnarray}
\zeta \equiv \frac{m_{\rm WD} + m_{\rm P}}{1m_{\odot}}.
\end{eqnarray}

\noindent{}Now denote orbital elements in $N$-body simulation coordinates with
the superscript ``N''.  Then

\begin{eqnarray}
a_{\rm P}^{\rm (N)} &=& a_{\rm P} \zeta^{1/3} 
,
\\
a_{\rm A}^{\rm (N)} &=& a_{\rm A} \zeta^{1/3} 
.
\end{eqnarray}

\noindent{}Other orbital elements such as eccentricity, argument of pericentre
and mean anomaly remain the same in both coordinate systems.  Also, importantly, time
is equivalent in both systems.  Consequently, the only transformation which needs
to take place is with the semimajor axes of both the planet and asteroid.

\subsection{Parameter choices}

The initial conditions for a $N$-body integrator require a mass, 
three positions and three velocities to be inserted for
each object.  In order to detect collisions with the star, the radius of the
star must also be specified.

We model evolution during the WD phase only, and hence adopt a fiducial WD mass of 
$m_{\rm WD} = 0.6m_{\odot}$.  In order to compare and provide a check on the results
from \cite{debetal2012} and \cite{frehan2014} (see Section 5.2), we 
give our planet a mass that is approximately
equal to Jupiter, such that $m_{\rm P} = 10^{-3}m_{\odot}$.  Consequently, our scaling
factor is $\zeta = 0.601$.  Regarding the radius of the star, we instead wish to use
the Roche (or disruption) radius of the star -- within which an asteroid would be 
tidally disrupted.  The WD's Roche radius is dependent on the properties of the 
asteroid that is being disrupted (see Section 2 of \citealt*{veretal2014b}).  Because we 
do not make assumptions about the internal properties of the asteroids, we simply adopt a
fictitious WD radius of $10^6$ km $\approx 1.4378 R_{\odot} \approx 0.006684$ au.  This value
conforms well to a conservatively large estimate of the WD's Roche radius.  Any 
asteroids which enter this radius are flagged as having collided with the WD. With regards to the planet's radius we used a radius which corresponds to a mass of 0.001 $m_{\odot}$ and a density of 1 $g/cm^3$ (so $R_{\rm P}=78,000$ km or 1.09 $R_{\rm Jupiter}$). 

Now consider the orbital elements of the planet and asteroid.
In the planar TBP, the inclinations and longitudes of ascending
nodes of the planet and asteroid can be and are set to zero without
loss of generality.  In the RTBP, because the asteroid is massless, 
astrocentric coordinates are equivalent
to Jacobi coordinates; the latter was used 
for periodic orbits.  In the CRTBP, because the planet is on a 
fixed circular orbit, $e_{\rm P}=0$ and
the choice of $\omega_{\rm P}$ is irrelevant; we fix it at
zero for all simulations.  The adopted 
value of $M_{\rm P}$ determines the planet's location along its
circular orbit.  We set $M_{\rm P} = 0^{\circ}$ initially in all
simulations in order to maintain the same reference geometry.

In all cases we adopt $a_{\rm P} = 10$ au, or $a_{\rm P}^{\rm (N)} = 8.439009789$ au.
The value of $a_{\rm P}$ is realistic because a relatively circular planet at a few 
to 5 au (like Jupiter) on the main sequence
will typically extend its orbit by a factor of 2-3 while maintaining its eccentricity 
\citep{veretal2011a} during the GB phases of a Solar-like or slightly more massive 
star \citep{dunlis1998,schcon2008,verwya2012}.  During the GB phases, a planet out
that far is unlikely to be engulfed due to star-planet 
tides \citep{kunetal2011,musvil2012,adablo2013,norspi2013,viletal2014,staetal2016}.

The four remaining parameters are then the ones we vary within our suite of
numerical simulations: $(a_{\rm A}^{\rm (N)}, e_{\rm A}, \omega_{\rm A}, M_{\rm A})$.
We describe and exhibit these choices in Section \ref{ing}, along with the results.

\subsection{Numerical integrator}

We perform our numerical simulations with the conservative
Bulirsch-Stoer integrator from the modified version of the {\sc mercury} 
\citep{chambers1999} integration suite that was used in \citep{vermus2013}.  We set the 
ejection radius at $3 \times 10^5$ au, which exceeds the Hill ellipsoid for a star
in the solar neighbourhood \citep{vereva2013}.  We adopt a highly accurate tolerance parameter 
of $10^{-13}$, which allows us to achieve conservation of energy and angular 
momentum to one part in $10^{-8} - 10^{-12}$. The Jacobi constant has typical variations of $10^{-4}$ to $10^{-3}$ (occasionally smaller), but no secular drift. Additionally, the computation of the Tisserand parameter \citep{bonwya2012} for the stable cases helps confirm our results.

The extent of our exploration of this region of phase space is limited by 
computational resources and integration timescales.
We evolve every one of our systems for  
the lifetime of the Universe (approximately 14 Gyr), corresponding to 
$\sim 1.1 \times 10^9$ orbits.  
Our output resolution is 1 Myr.

\section{N-body integration results}  \label{ing}

In order to explore the 4-dimensional ($a_{\rm A}, e_{\rm A}, \omega_{\rm A}, M_{\rm A}$) 
phase space, we vary selected pairs of variables in the vicinity of the 
$2$:$1$ periodic orbits of the planar CRTBP.  Our primary interest is connecting
the properties of $N$-body simulation outcomes with $(a_{\rm A}, e_{\rm A})$ 
portraits such as those in Fig. \ref{mapsea} (subsection \ref{aepor}) and $(M_{\rm A}, \omega_{\rm A})$ 
portraits such as those in Fig. \ref{mapsapowMMOD} (subsection \ref{Mompor}).
First, however, we comment on some global properties of our simulations
(subsections \ref{instime} and \ref{orbvar}).  We finish this section
with some specific examples of system orbital evolution (subsection \ref{indruns}).

\subsection{Instability timescales}  \label{instime}

The most important result relating to WD systems regards
instability timescales.  Across all of our 2847 simulations in the planar CRTBP -- including
858 simulations with a stable asteroid that were run for the age of the Universe (14 Gyr) -- an 
asteroid collision with the planet or WD never occurred after 36 Myr.  Further, a collision
occurred after 1 Myr only 17 times.  Further still, many of our simulations, as illustrated
below, specifically targeted the regions around unstable periodic orbits.
The two main consequences of this finding are (1) the planar CRTBP $2$:$1$ architecture 
cannot produce long-timescale ($\gtrsim 0.3$\% of the age of the Universe) collisions, 
and (2) that this architecture cannot explain
observed rates of WD pollution, in line with the findings of \citet{frehan2014}.
We will discuss these points more in Section \ref{DiscSec}.

\begin{figure}
\centerline{
\includegraphics[height=5.5cm]{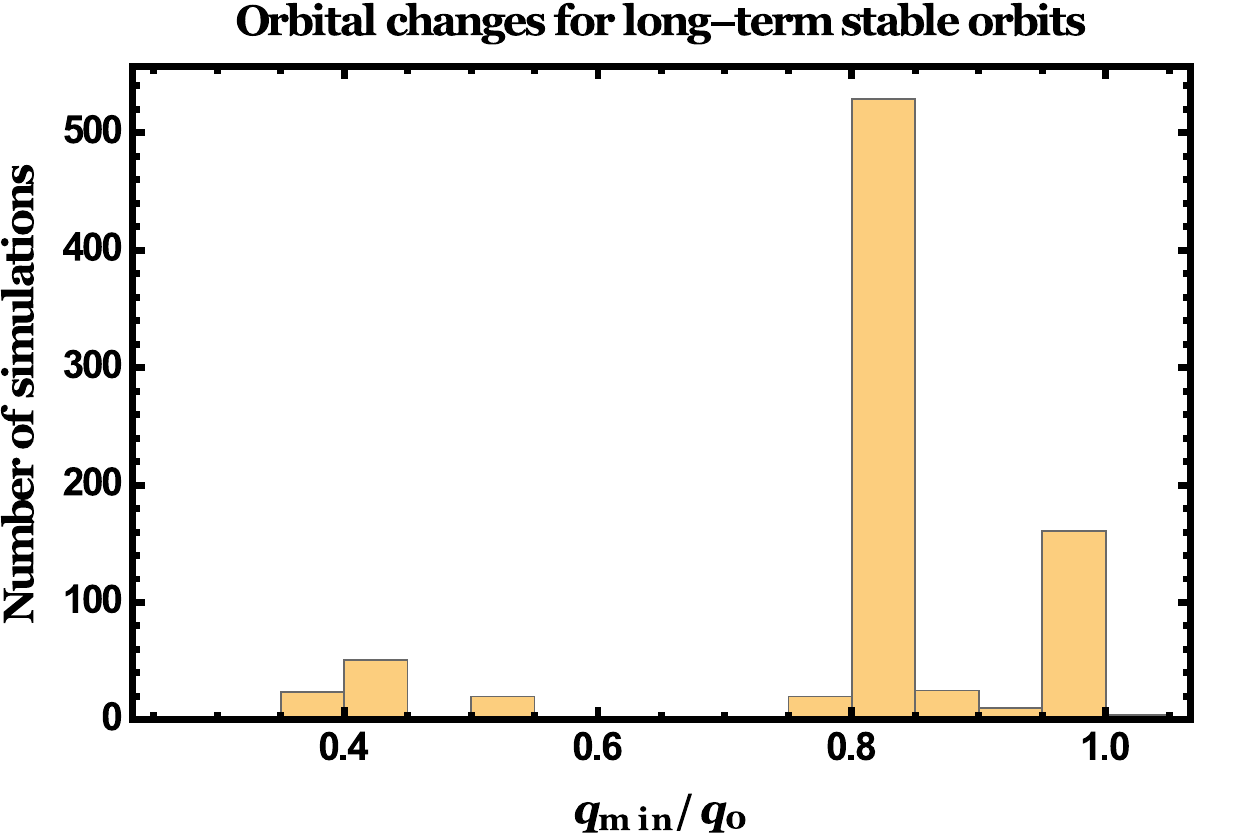}
}
\caption{
Number of stable systems integrated over 14 Gyr which experience a given 
maximum change in pericentre distance.  Although this distribution is 
likely reflective of our initial condition choices, the lowest value of about 0.4
is more importantly probably reflective of a global minimum.
}
\label{Perichange}
\end{figure}

\subsection{Orbital variations}  \label{orbvar}

For studies relevant to WD pollution, a particularly important orbital
quantity to measure is how the pericentre distance $q$ changes with time.
Let $q_0$ represent the initial pericentre distance and $q_{\rm min}$
represent the minimum value of $q$ achieved throughout the simulation.
We plot $q_{\rm min}/q_0$ in Fig. \ref{Perichange} for all simulations
which remained stable and ran for the age of the Universe.

This distribution is naturally highly-dependent on the initial conditions we chose for our
simulations. The peak around $q_{\rm min}/q_0 \approx 0.8$ might simply reflect this dependency.
However, because we deliberately attempted to sample key {\it unstable} regions
of phase space, the minimum ratio attained of $q_{\rm min}/q_0 \approx 0.4$ is likely
to be representative of a global minimum.  At least, achieving a value less than about 0.4
is highly unlikely (at the $0.1$\% level).  A value of 0.4 does not allow the asteroid
to get remotely close to the Roche radius of the WD unless the initial pericentre
is set to within about $0.02$ au, which is only possible due to a scattering event
after the parent star has become a WD.

\subsection{Varying pairs of $a_{\rm A}$ and $e_{\rm A}$}  \label{aepor}

\begin{figure}
\centerline{
\includegraphics[height=8.2cm]{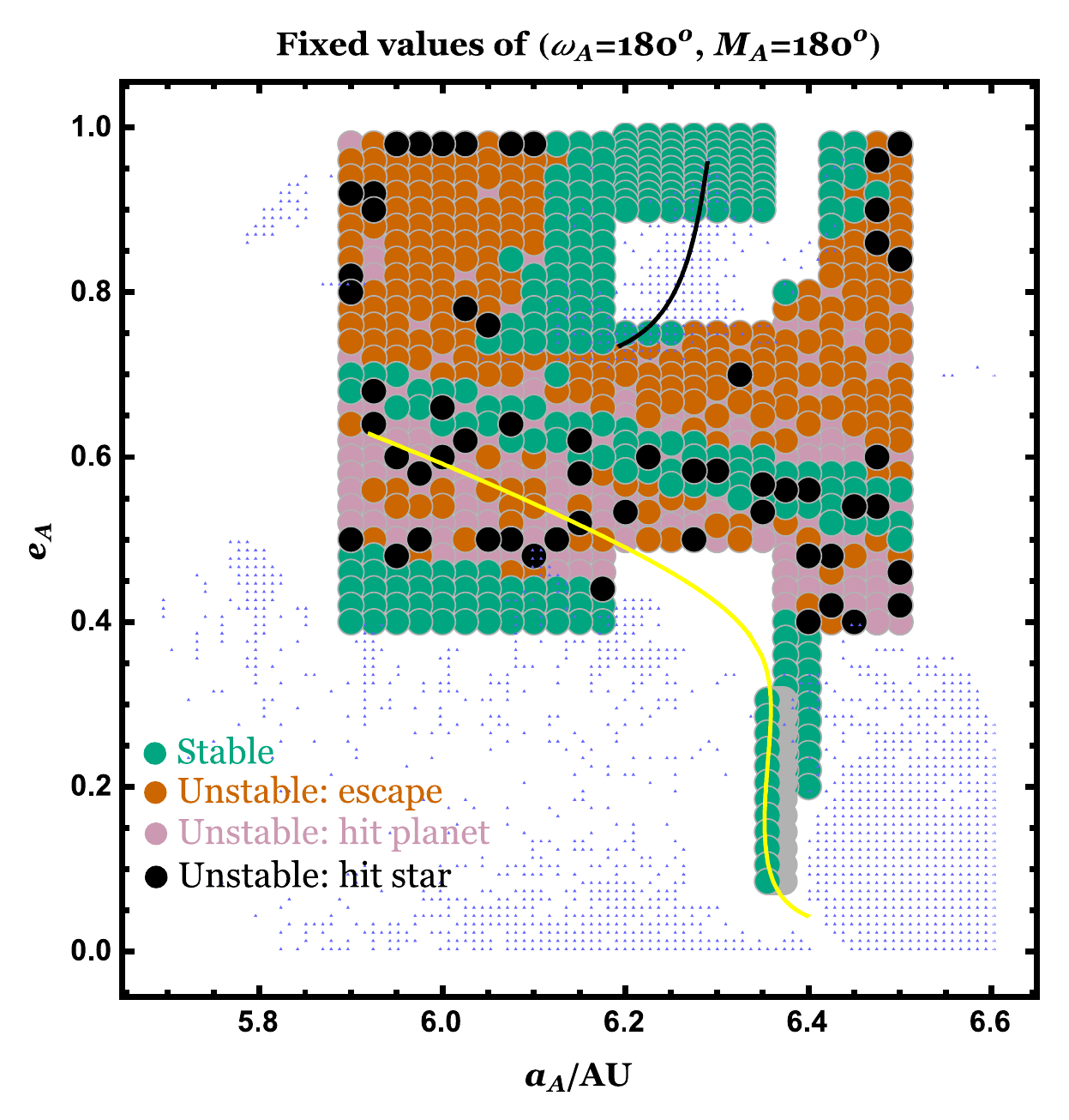}
}
\centerline{}
\centerline{
\includegraphics[height=8.2cm]{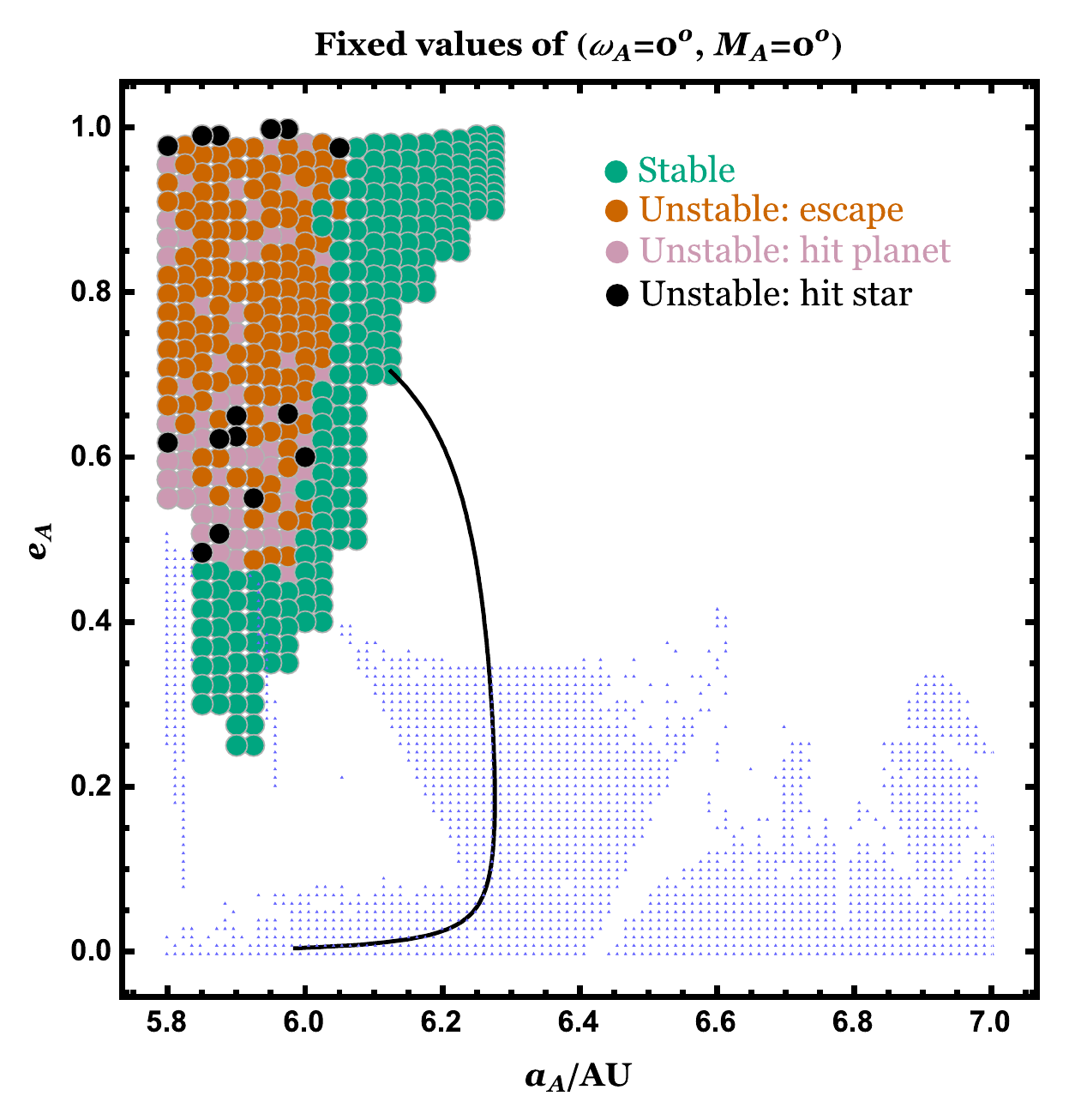}
}
\caption{
Stability portrait from 14 Gyr $N$-body simulations for fixed $(\omega_{\rm A},M_{\rm A})$.
Green spheres indicate stable systems and other dots indicate unstable systems.
Orange, purple and black spheres respectively refer to systems where the asteroid
escaped, hit the planet, and hit the star.  The gray-looking spheres are the
result of densely packed green spheres.  These plots should be compared with
Fig. \ref{mapsea}; Families of stable (black) and unstable (yellow) periodic orbits, as well as grid points where $log(DFLI)\leq 1.2$ (pale blue triangular symbols) are overplotted.
}
\label{Nbodyae}
\end{figure}

Now we directly compare Fig. \ref{mapsea} with similarly-scaled
figures that display the results of our long-term integrations
(Fig. \ref{Nbodyae}).  The top and bottom panels of Fig. \ref{Nbodyae}
correspond to the $\omega_{\rm A} = M_{\rm A} = 180^{\circ}$ and 
$\omega_{\rm A} = M_{\rm A} = 0^{\circ}$ cases respectively, like the 
left and right panels of Fig. \ref{mapsea}.  The outcomes of the $N$-body
integrations are parametrized with different colours, indicating stability
(green), escape (orange), collision with the planet (purple) and collision
with the star (black). Hence, stability as defined in the N-body simulations does not reflect the long-term stability in the vicinity of stable periodic orbits where the motion is regular, but solely the absence of ejection, collision with the star, or collision with the planet. As we have already mentioned in Section \ref{po}, chaotic orbits are detected in the neighbourhood of unstable periodic orbits. In some cases, despite the irregularity in the evolution, the initial orbital elements do not change significantly over time and thus, these orbits can be considered as ``stable'' (see e.g. the examples in Section \ref{indruns}).

Hence, our comparisons are inexact: Fig. \ref{mapsea} displays the 
periodic orbits as well as the DFLI maps, whereas Fig. \ref{Nbodyae} 
strictly illustrates qualitative system outcomes.  In this figure we observe

\begin{enumerate}
 \item{Distinct blocks of stable systems. The locations of several of these blocks 
correspond to both regions which are close to stable periodic orbits and to regions 
containing DFLI regular orbits.  One diagonal strip of stable systems, however, on the
upper panel of Fig. \ref{Nbodyae} with $0.5 \le e_{\rm A} \le 0.7$, is
not a solid block, but instead is punctuated with collisional instability.
A solid stable block also encompasses the lower
tail ($e_{\rm A} \gtrsim 0.4$) of the unstable periodic orbit family 
(the gray dots represent densely-packed green dots).  In this tail, the DLFI
includes a mixture of regular and chaotic orbits and does not tend strongly
towards either type.
}
 \item{Distinct blocks of unstable systems. These blocks do encompass DFLI chaotic
orbit regions.  One unstable block surrounds the upper portion ($e_{\rm A} \gtrsim 0.4$) of the
unstable periodic orbit family.  This block features a mix of all types of instability.
}
 \item{An inhomogeneous distribution of incidents of star-asteroid and star-planet collisions
throughout the phase space.  Escape is the dominant form of instability in these systems, and
appear in blocks.  For the WD case, all collisions occur on too short of a timescale (under 36 Myr)
to explain Gyr pollution timescales, regardless of the collision distribution in phase space.}
\end{enumerate}

Overall, the correspondence between periodic orbits, DLFI maps, and long-term 
$N$-body simulations is robust. The low-eccentricity tail of the family of unstable periodic orbits is the most intriguing discrepancy, at first glance. Poincar\'e surface of sections have on the one hand, revealed that the chaotic regions are bounded and hence, escape cannot occur. On the other hand, they showed that such regions are sensitive to Jupiter's eccentricity \citep[see e.g.][]{mifm95,NeFe97,havo01}. Furthermore, the results of N-body simulations shown in Fig. \ref{Nbodyae} and depicting the existence of islands of stability within chaotic regions as $a_{\rm A}$ and $e_{\rm A}$ vary in each configuration (i.e. when $\omega_{\rm A} = 0^{\circ}$ or $\omega_{\rm A} = 180^{\circ}$) are in good agreement with the results presented with surfaces of section in \citet{MoMo93}.

\subsection{Varying pairs of $M_{\rm A}$ and $\omega_{\rm A}$}  \label{Mompor}

Now let us explore this correspondence with the DFLI maps on plane $(M_{\rm A},\omega_{\rm A})$ from Fig. \ref{mapsapowMMOD}. These phase portraits illustrate
how stability changes as $M_{\rm A}$ and $\omega_{\rm A}$ are varied, instead of
$a_{\rm A}$ and $e_{\rm A}$.  These portraits do intersect with both unstable
and stable periodic orbits at the points given in that figure caption, but essentially
focus on the regions surrounding those orbits. The regions of greatest interest
to us are the yellow strips, which are filled with irregular orbits that {\it may}
become unstable on the timescales we are interested in (Gyr).

\begin{figure}
\centerline{
\includegraphics[height=6.5cm]{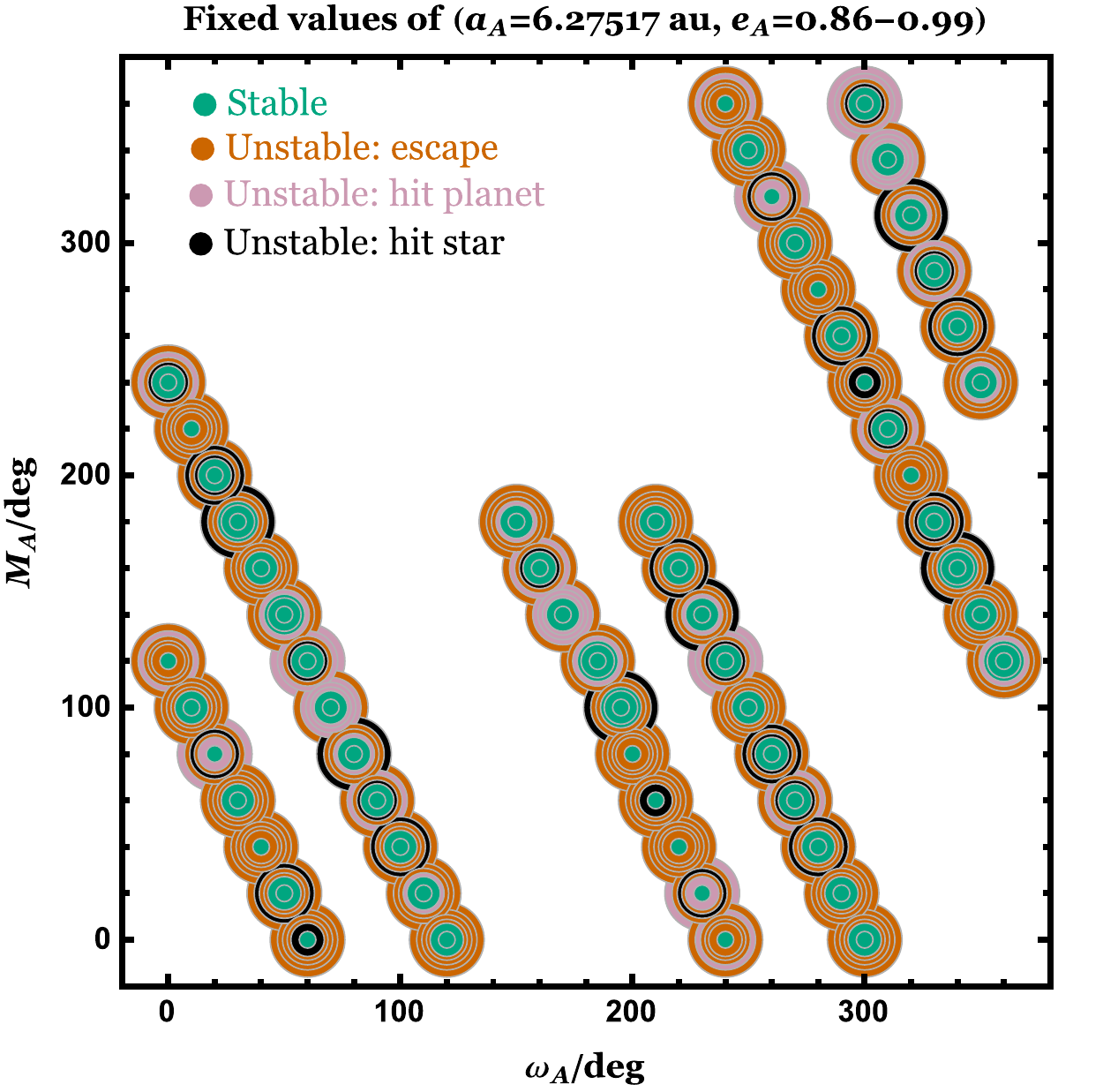}
}
\centerline{}
\centerline{
\includegraphics[height=6.5cm]{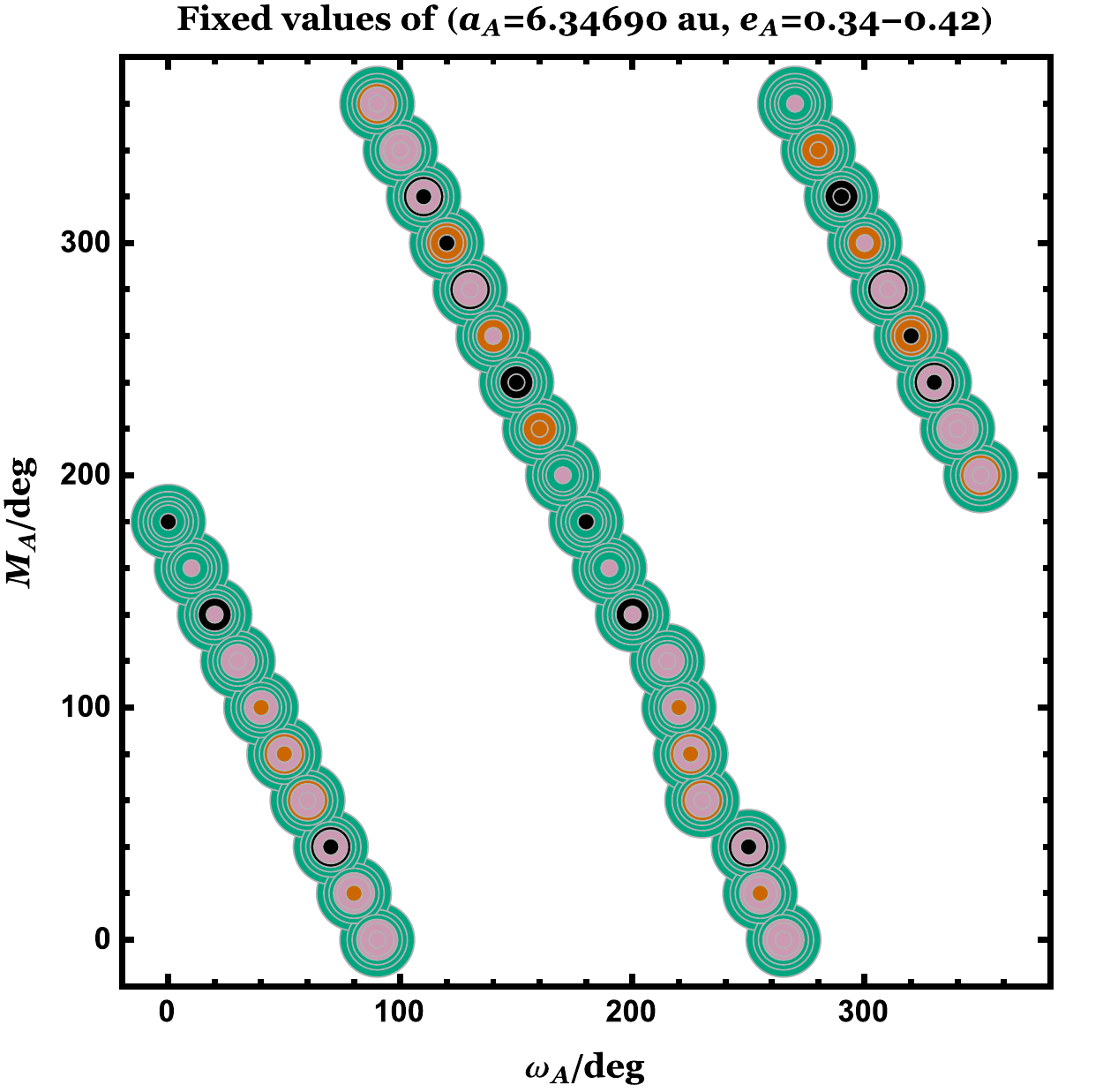}
}
\centerline{}
\centerline{
\includegraphics[height=6.5cm]{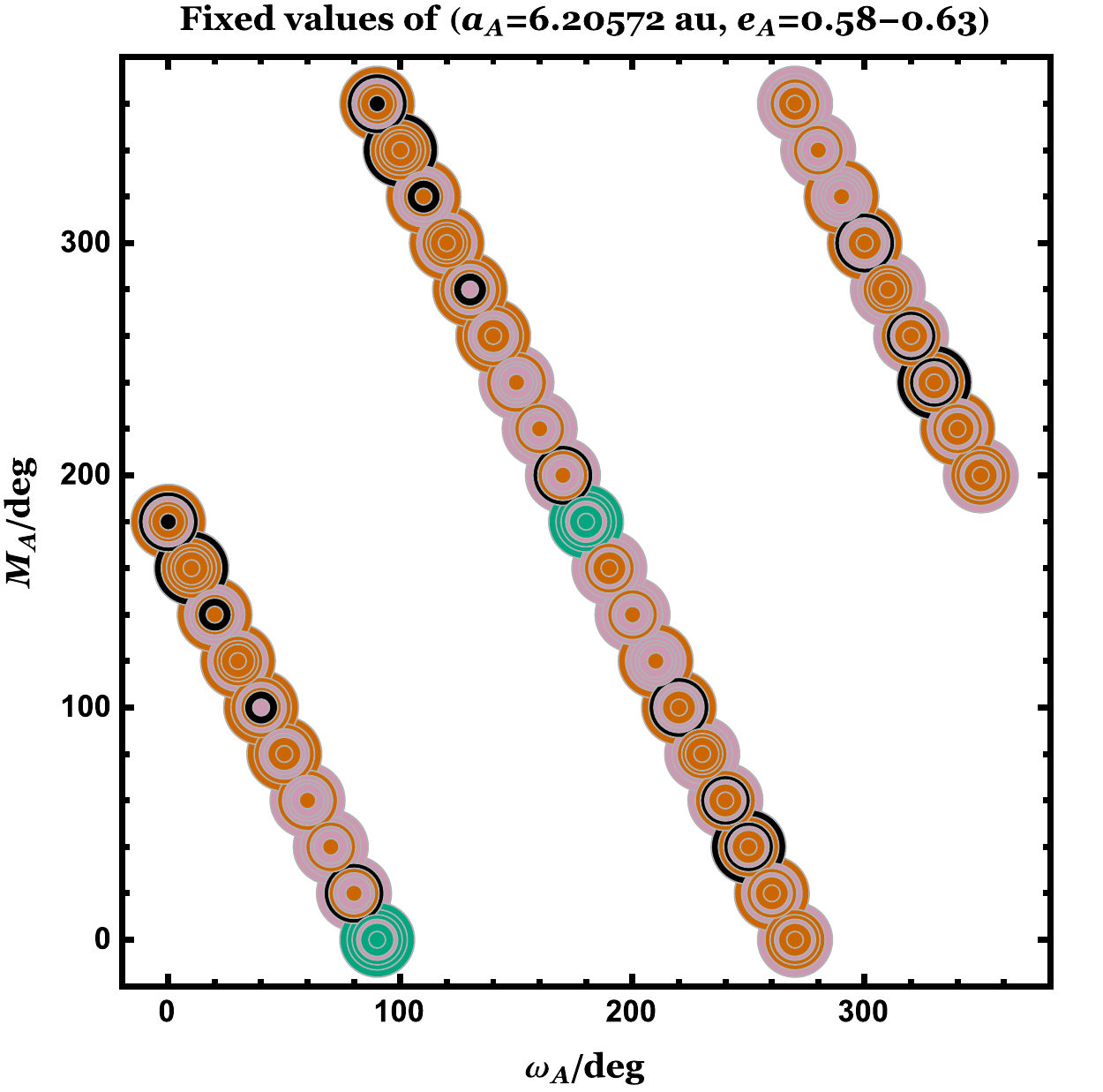}
}
\caption{
Stability portrait from 14 Gyr $N$-body simulations for fixed $a_{\rm A}$.
Dots of different sizes refer to fixed values of $e_{\rm A}$.  The
largest to smallest dots refer to ({\it top panel}) 
$e_{\rm A}= \left\lbrace 0.86,0.88,0.90,0.95,0.97,0.99 \right\rbrace$,
({\it middle panel}) 
$e_{\rm A}= \left\lbrace 0.34,0.35,0.36,0.38,0.40,0.42 \right\rbrace$
and ({\it lower panel})
$e_{\rm A}= \left\lbrace 0.58,0.59,0.60,0.61,0.62,0.63 \right\rbrace$
Colours are as in Fig. \ref{Nbodyae}.
}
\label{NbodyomM}
\end{figure}

We present the results of the $N$-body simulations in Fig. \ref{NbodyomM}.
Note that we choose initial conditions for the simulations which roughly correspond
to all of the yellow strips in Fig. \ref{mapsapowMMOD}.  For each point in each plot in 
Fig. \ref{NbodyomM}, we ran multiple simulations
at different values of $e_{\rm A}$ (provided in the caption).  We observe

\begin{enumerate}
 \item{Variations in $e_{\rm A}$ on the order of hundredths often completely changes the
outcome of the simulations.  In the top plot, at a given point, as $e_{\rm A}$ increases,
unstable systems give way to stable systems, because the island of stability is reached (compare Figs. \ref{mapsea}a, \ref{mapsapowMMOD}a and \ref{mapsapoperweMOD}).  These unstable systems predominately feature
escape.  In the middle plot, at a given point, as $e_{\rm A}$ increases,
stable systems give way to unstable systems (which are a mix of escape and collisions), because the chaotic region for the semimajor axis that is fixed begins roughly when $e_{\rm A}>0.4$ (compare Figs. \ref{mapsea}a, \ref{mapsapowMMOD}b and \ref{mapsapoperweMOD}). 
In the bottom plot, nearly all the simulations become unstable.  However, the character
of the instability changes as one varies $e_{\rm A}$.
At no point in any of the three plots are all concentric circles the same colour.
}
 \item{For a fixed value of $e_{\rm A}$, as one moves along the strips, the outcome of the
simulations (stable or unstable) remains largely the same.  This trend is apparent in all three plots.
}
 \item{Collisions between the asteroid and star are inhomogeneously and infrequently distributed.
In no set of concentric circles are more than two coloured black.
}
 \item{The two (they are equivalent) exceptions in the bottom plot feature green concentric circles of nearly all stable
simulations. They are linked with the outcome that is also seen in the top panel of Fig. \ref{Nbodyae} for this specific range in the eccentricities and the fixed value of the semimajor axis and justified therein accordingly (compare Figs. \ref{mapsea}a, \ref{mapsapowMMOD}c and \ref{mapsapoperweMOD}). Also, note how the frequency of green circles decreases as one strays from these equivalent configurations. 
}
\end{enumerate}

Overall, Figs. \ref{mapsapowMMOD} and \ref{NbodyomM} demonstrate the importance of the initial
orbital angles of the asteroid when determining long-term stability. The role of eccentricity
is crucial, and demonstrates predictive power with clear trends.

\subsection{Specific evolution examples}  \label{indruns}

Even within the confines of the $2$:$1$ planar CRTBP, the asteroid's
osculating orbit may remain nearly static or vary dramatically depending
on its initial conditions.  Consequently, there is no one representative
orbit.  Nevertheless, in this subsection we provide a few examples of
asteroid evolution in order to at least impart to the reader a flavour of 
the variety seen across the simulations.

\begin{figure}
\centerline{
\includegraphics[height=6.8cm]{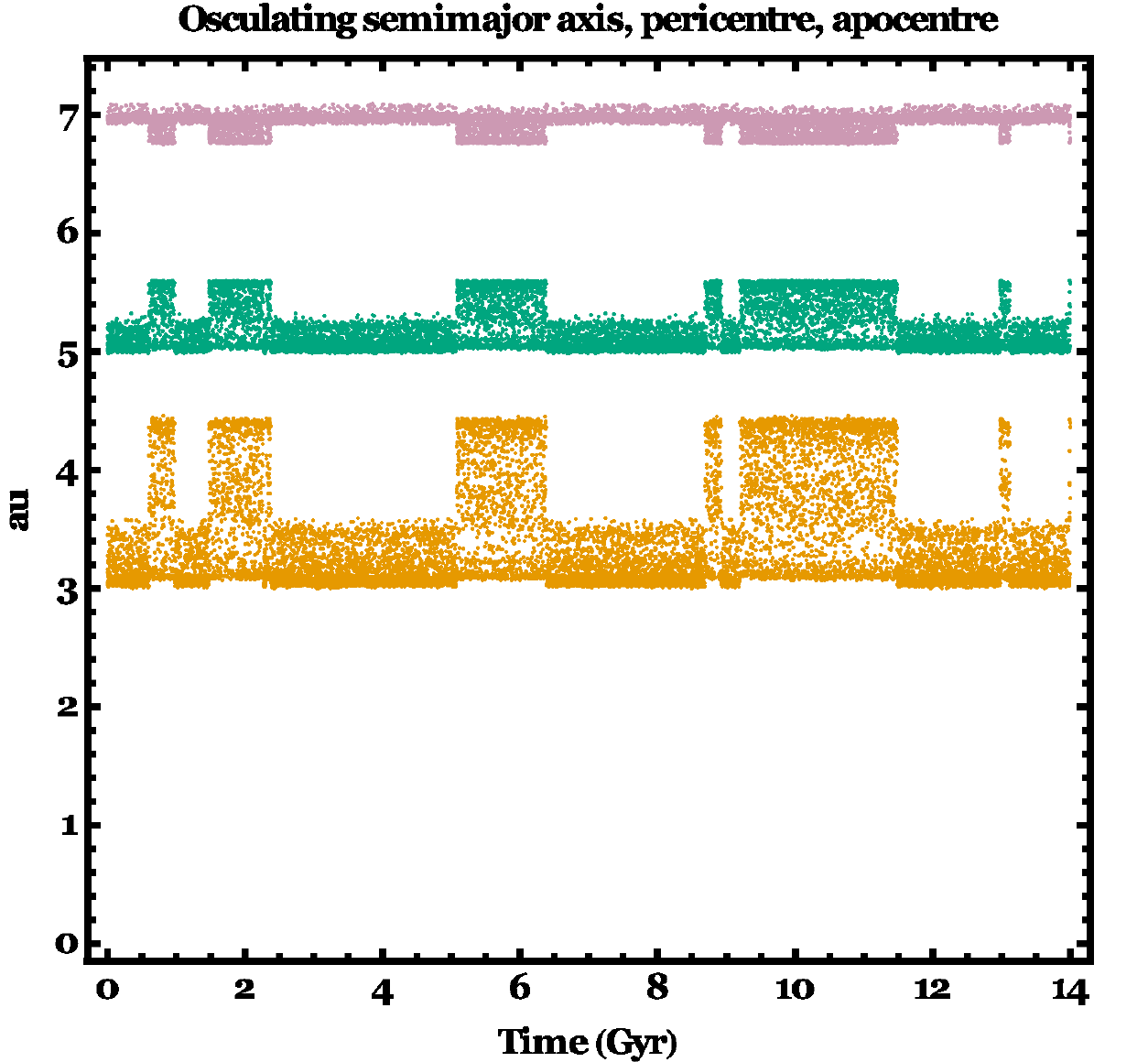}
}
\centerline{
\includegraphics[height=6.8cm]{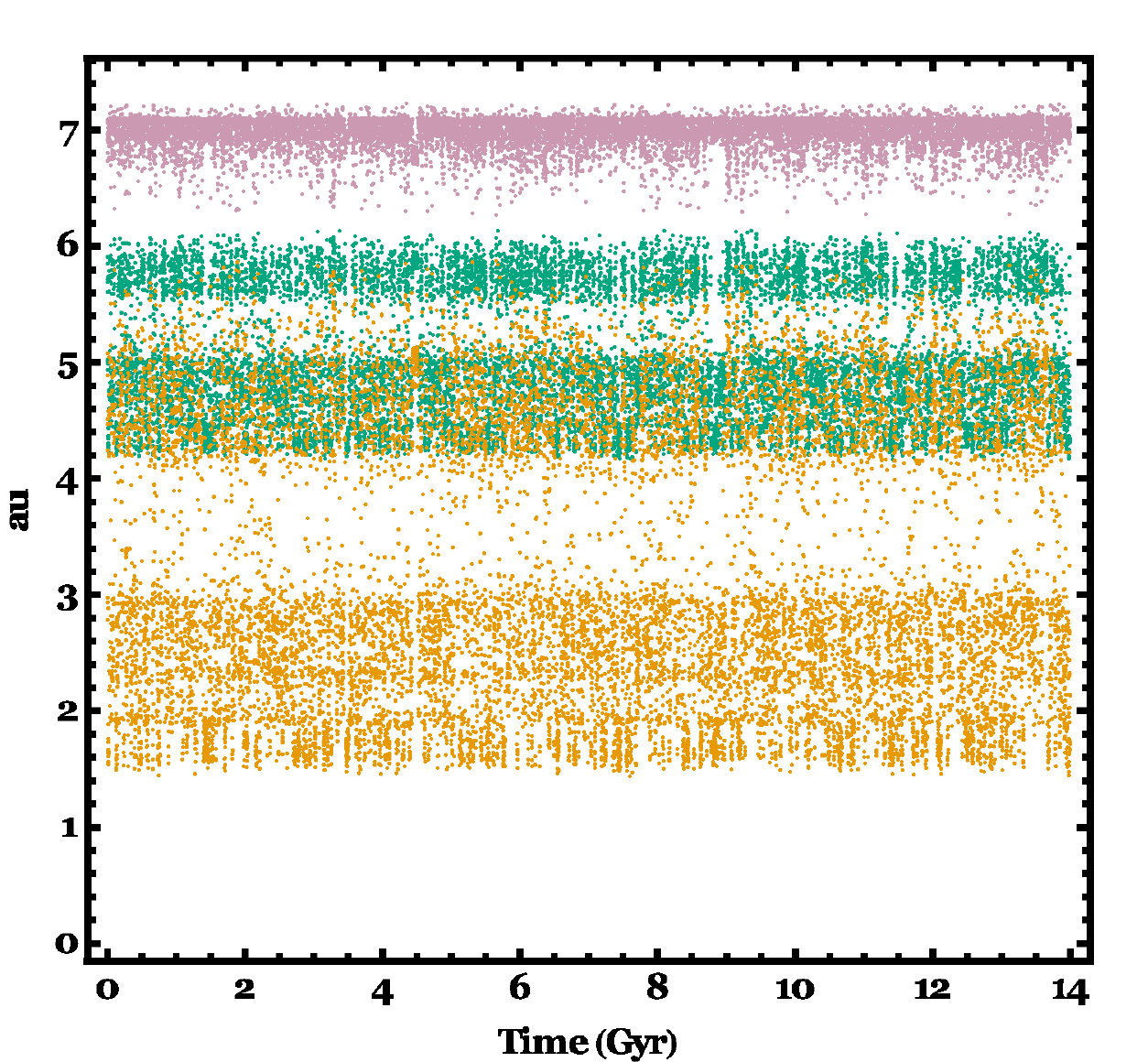}
}
\centerline{
\includegraphics[height=6.8cm]{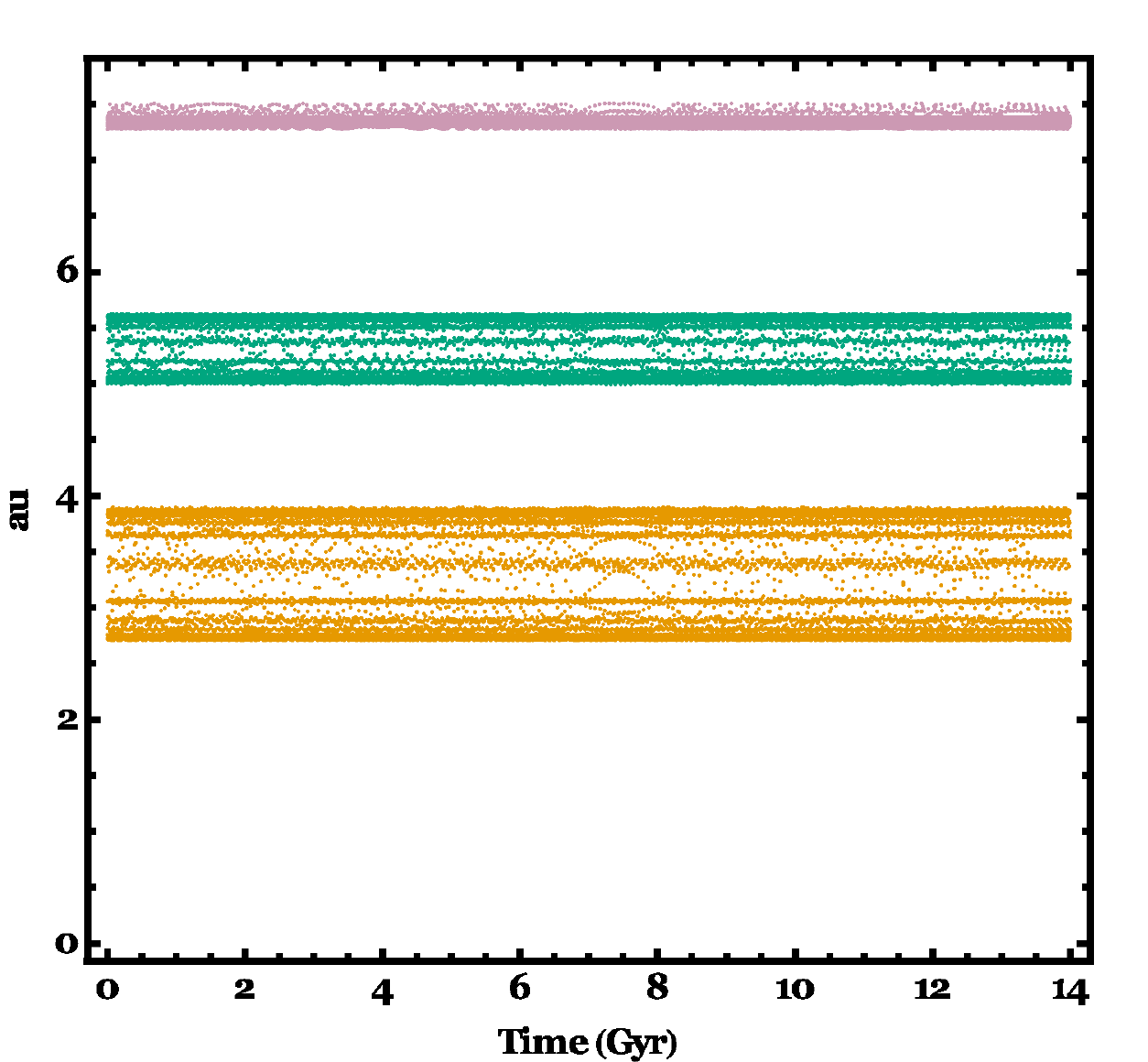}
}
\caption{
The evolution of the semimajor axis (green), pericentre (orange) 
and apocentre (purple) of an asteroid in three individual systems.
In all three systems, $a_{\rm A} = 5.95$ au, $\omega_{\rm A}=0.0^{\circ}$ 
and $M_{\rm A}=0.0^{\circ}$.  In the top, middle and bottom panels,
respectively, $e_{\rm A} = \left\lbrace 0.35, 0.4148, 0.458\right\rbrace$.
The plots illustrate different qualitative orbital behaviour by changing
the eccentricity by about 0.1.  Note that the pericentre experiences the
greatest variations for the middle eccentricity value.
}
\label{Indevo1}
\end{figure}

First consider Fig. \ref{Indevo1}, which exhibits
the evolution for three simulations with identical
initial orbital parameters (given in the caption) except
for $e_{\rm A}$.  As $e_{\rm A}$ increases from
0.35 to 0.458 across the three simulations, the qualitative
behaviour of the osculating orbit changes in unpredictable ways.
The pericentre variation is greatest for the middle panel
(from about 1.5 au to 6.0 au).  In the top panel, the orbit
oscillates back and forth between two different modes of evolution.
In the bottom panel, the semimajor axis and pericentre appear to be
rigidly bound within differently-sized ranges.  When evaluating
these plots, please recall that the output resolution for our
simulations was 1 Myr.

\begin{figure*}
\centerline{
\includegraphics[height=6.8cm]{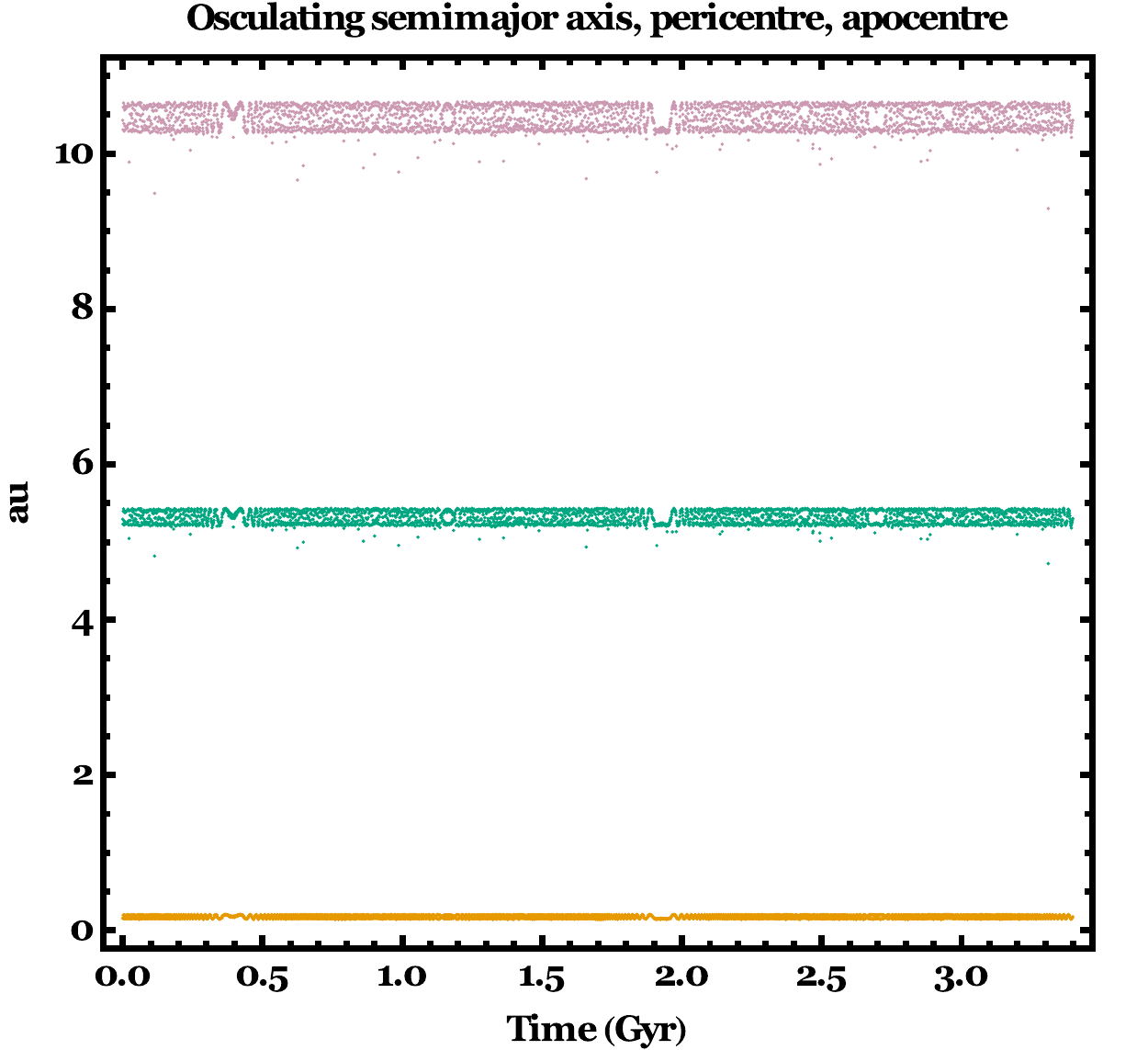}
\includegraphics[height=6.8cm]{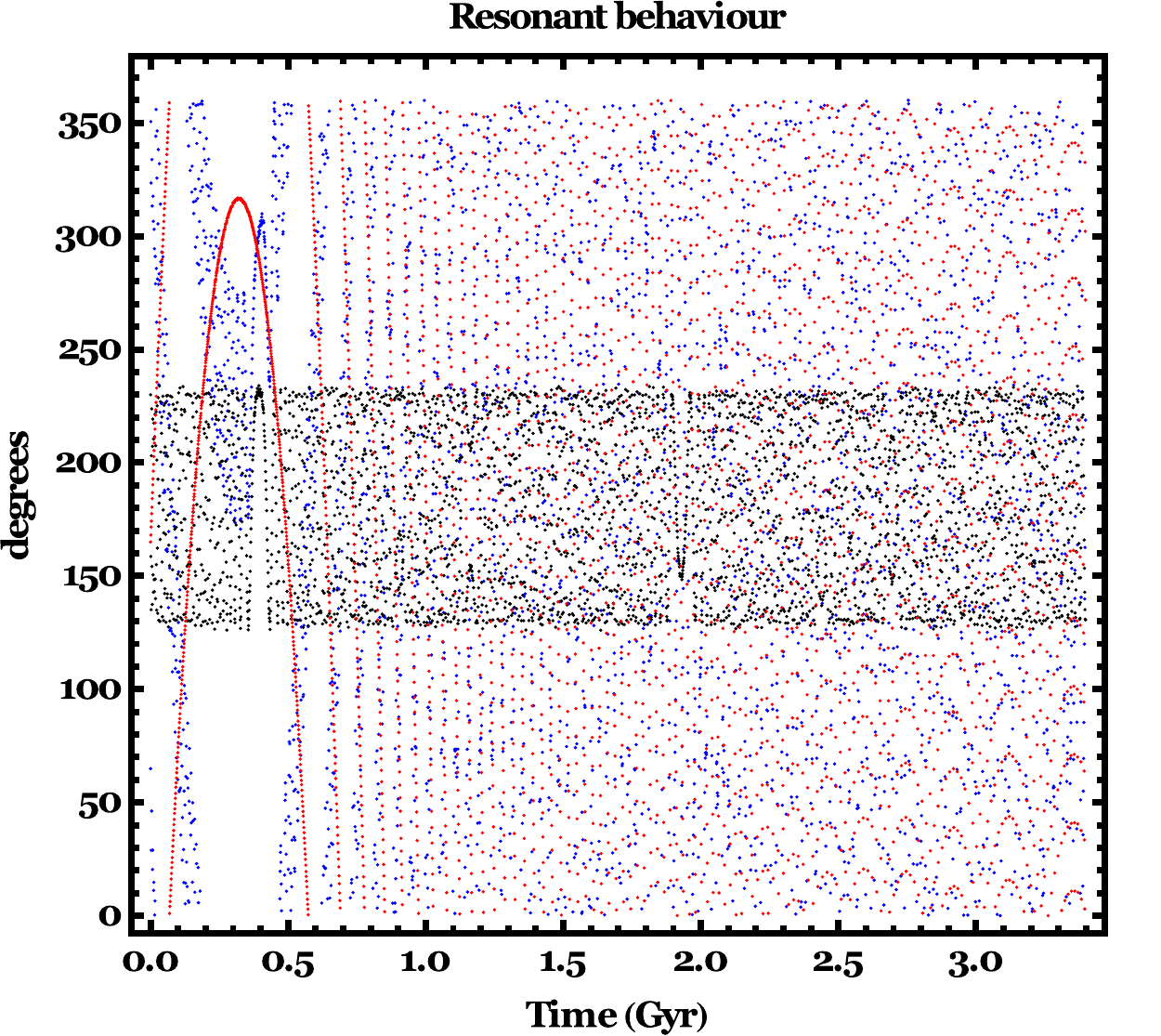}
}
\caption{
The evolution of an initially and otherwise highly eccentric system
($e_{\rm A} = 0.97$) system which exhibits some resonant behaviour.
Other initial parameters are $a_{\rm A}=6.2517$ au, $\omega_{\rm A} = 195^{\circ}$
and $M_{\rm A}=100^{\circ}$.  The left panel illustrates the evolution of the
semimajor axis (green), pericentre (orange) and apocentre (purple), and the right panel illustrates
the evolution of the two resonant $2$:$1$ angles (black and blue) and
the apsidal angle (red).  The black points, which correspond to the resonant
angle which includes the asteroid's argument of pericentre, appears to librate about $180^{\circ}$
throughout the 3.5 Gyr of evolution shown, and shows two distinctive features at about 0.4 Gyr
and 1.9 Gyr.  The circulating apsidal angle is slowest at about 0.3 Gyr.
}
\label{Indevo2}
\end{figure*}

Figure \ref{Indevo2} illustrates a more extreme case, where
the asteroid is initially highly eccentric ($e_{\rm A} = 0.97$).
For this system, we show the evolution of the two $2$:$1$ MMR
angles in the right panel (black and blue), along with the evolution of the
apsidal angle (red).  The black dots clearly librate about $180^{\circ}$
with an amplitude of nearly $50^{\circ}$. The semimajor axis, pericentre and apocentre
appear to change little, although the stray purple and green dots
below the main stripes indicate brief but potentially dramatic
swings from the initial orbit.  Inhomogeneities in the stripes
themselves (at about 0.4 Gyr and 1.9 Gyr) show up in the right panel
as very slow oscillations in the black curves.  Alternatively, the 
dramatically slow circulatory motion indicated by the blue and red curves 
at about 0.3 Gyr does not appear to manifest itself in the left panel. In order to test if these features are a result of aliasing, we have run 5 additional simulations, each for 7 Gyr. In these simulations, the original output intervals were decreased by factors of 10, 7, 5, 3 and 2. The results demonstrate that the resonance in Fig. \ref{Indevo2} appears to be real. In all cases, the angle librates and never deviates from the above-mentioned range.

\section{Discussion} \label{DiscSec}

In order to place our findings in the proper context, we consider here how they relate to real planetary systems.  Partly due to history and to the voluminous and accurate set of available Solar system asteroid data, the vast majority of literature on long-term asteroid evolution is Solar System-specific \citep[e.g.][]{botetal2000,morbidelli2002b,obrsyk2011}.  Inherent in Solar system studies are assumptions about orbital architectures and radiative physics which do not necessarily apply in extrasolar systems, especially post-main-sequence exosystems.

\subsection{Comparison with Solar system studies}

The Solar system is different from the planar CRTBP because the former contains more than one planet, and all of the Solar system planets have eccentric orbits which are at least slightly noncoplanar with asteroids and the Sun.  Consequently, collisions with the Sun can occur through scattering interactions with multiple planets, leading to a much higher fraction of collisions than in our case \citep{faretal1994,glaetal1997,minmal2010}.  The presence of multiple planets does not, however, preclude the possibility of long-lived (on Gyr timescale) stable ``islands" where asteroids can reside in or close to the $2$:$1$ mean motion resonance with Jupiter \citep{chretal2015}.  Asteroids in such islands may also be shaped by radiation-based perturbations such as the Yarkovsky effect \citep{broetal2005}, which can affect the long-term evolution \citep{brovok2008} in ways that are missed in the the planar CRTBP studied herein.

\subsection{Comparison with WD system studies}

In contrast, extrasolar systems offer a wide variety of architectures,
The majority of all known exosystems contain one exoplanet, leaving
open the possibility of the RTBP being widely applicable outside
of our Solar system.  Individual asteroids are currently undetectable, 
except in WD systems, where their tidally-shorn innards are observed orbiting WDs 
or accreting onto them.

Asteroids, comets and planets which orbit stars that leave
the main sequence become subject to an assortment of 
important physical processes whose previous influence was
negligible:

\begin{enumerate}

\item Stellar mass loss, which causes an outward expansion and possible stretching and breaking
of a bound osculating orbit 
\citep{omarov1962,hadjidemetriou1963,veretal2011a,adaetal2013,veretal2013a,voyetal2013}

\item Stellar radius enlargement, leading to engulfment or tidal disruption
\citep{villiv2009,kunetal2011,musvil2012,adablo2013,norspi2013,viletal2014,staetal2016},

\item Stellar wind drag, which usually acts in opposition to mass loss 
\citep{bonwya2010,donetal2010,veretal2015a}

\item Variations in stellar luminosity, which lead to YORP-induced fission \citep{veretal2014a},
Yarkovsky-induced dispersion \citep{veretal2015a}, inward drag from radiation pressure and 
Poynting-Robertson effects \citep{bonwya2010,donetal2010,veretal2015b}
and sublimation \citep{veretal2015c}

\end{enumerate}

These effects redistribute surviving bodies and debris in complex 
and still poorly-understood ways.  Regardless, observations reveal
that this material somehow is frequently and significantly perturbed 
close to, and ultimately onto, the resulting WD.  In fact
planetary remnants are directly detected in the atmospheres of 
between one quarter and one half of all Milky Way WDs 
\citep{zucetal2003,zucetal2010,koeetal2014}.  These signatures are
referred to as {\it pollution} and are thought to arise from accretion 
from compact discs \citep{rafikov2011a,rafikov2011b,metetal2012,rafgar2012,
wyaetal2014} which are
readily observed \citep[e.g.][]{beretal2014,rocetal2014,wiletal2014,radetal2015} 
and are thought to have 
formed by the tidal disruption of incoming planetary bodies 
\citep{graetal1990,jura2003,debetal2012,beasok2013,veretal2014b}, such as
those disintegrating around WD 1145+017 \citep{vanetal2015}. WD pollution 
reveals the chemical composition of the accreeted
bodies \citep{zucetal2007,gaeetal2012,faretal2013,juryou2014,xuetal2014} and consequently
provides unambiguous bulk density data about exoplanetary interiors.

Asteroids which pollute WDs must be on highly eccentric orbits because
in order to avoid engulfment during the giant branch (GB) stages of stellar evolution,
the semimajor axis must be at least a few au.
In order to be perturbed into highly eccentric orbits, asteroids need instigators.
Three investigations \citep{bonetal2011,debetal2012,frehan2014} have dynamically modelled 
the interaction of asteroids with one planet during the GB phases of evolution, 
as well as up to 1 Gyr during the WD phase. \citet{bonver2015} considered the consequences
of a perturbation by a wide orbit stellar companion, and \cite{payetal2016a,payetal2016b}
determined how liberated moons would contribute to the dynamical architecture.

However, none of those 
studies included simulations which incorporated the Yarkovsky effect nor stellar wind drag
(although see Section 5.1 of \citealt*{frehan2014}), which 
could represent the dominant drivers of asteroid evolution during the GB phase \citep{veretal2015a}.
Therefore, how asteroids evolve during GB evolution, supposing they survive rotational spin-up 
\citep{veretal2014a}, remains an open question, one that we do not address here. 

The RTBP is particularly relevant to these three studies.
\citet{debetal2012} suggested that within the RTBP the eccentricity of an asteroid 
which resides near or inside of an MMR with a planet can gradually increase until the
asteroid achieves a 
WD-grazing orbit.  They performed
simulations with one planet on an eccentric orbit ($e_{\rm P} = 0.0488$) and with 
asteroids that were inclined with respect to the star-planet plane (J. Debes, 
private communication) in order to demonstrate their theory.  \citet{frehan2014} 
later explicitly explored how
the eccentricity of the planet affects collision rates with the WD when the
asteroid was confined to within $0.5^{\circ}$ of the WD-planet plane.  
They found that $e_{\rm P} > 0.02$
is a necessary condition to pollute WDs with a sufficiently high amount of material
to extrapolate to observations.

Although the numerical simulations of both studies ran for up to only 1 Gyr, our results affirm 
theirs and stringently rule out dynamical pollution mechanisms with
a single planet on a circular orbit at the $2$:$1$ commensurability.  The small inclinations 
adopted by \cite{frehan2014} 
suggests that the driver for collisions is planetary eccentricity rather than non-coplanarity.
Regardless, undertaking a systematic study for each of these orbital parameters 
-- as well as different commensurabilities -- is now required in order to pinpoint 
the most likely physical origin of long-timescale collisions with the WD.  Periodic orbits
represent an effective tool for this purpose, especially since so many families have already
been computed.

\section{Conclusion}

We have evaluated the predictive power of periodic orbits and their phase space surroundings to determine the long-term (1-14 Gyr) stability of systems containing one star, one planet and one asteroid in the planar CRTBP.  We utilized the three families of periodic orbits which are associated with the $2$:$1$ MMR, such that the asteroid is interior to the planet. Our resource-consuming 14~Gyr simulations revealed good agreement between the final dynamical outcome and the structure of phase space about the periodic orbits. Consequently, researchers can use these orbits as broadly reliable guides to choose sets of initial conditions ($a_{\rm A}^{(\rm N)}, e_{\rm A}, \omega_{\rm A}, M_{\rm A}$) for $N$-body simulations to predict whether a given system will remain stable over the age of the Universe. 

Of particular interest for Gyr-old WDs which host circumstellar debris discs and atmospheric metal pollution are \textit{unstable} planetary systems that feature very late collisions between the asteroid and star.  Therefore, in this paper we have focused on unstable cases.  Despite this emphasis, in none of our simulations did a collision occur after 36 Myr, thereby providing analytical backing to the findings of \citet{frehan2014} that a circular planet fails to pollute WDs with a sufficient number of asteroids at sufficiently late ages.

\section*{Acknowledgments}

We would like to thank the reviewer, A. Mustill, for his comments with regards to the figures, whose implementation enhanced our paper. We thank John H. Debes and Matthew J. Holman for useful discussions.
DV benefited by support by the European Union through ERC grant number 320964.

\bsp	

\label{lastpage}

\end{document}